\documentclass[letterpaper,11pt]{article}

\usepackage[top=1.0in,bottom=1.0in,left=1.0in,right=1.0in]{geometry}

\usepackage{amssymb}
\usepackage{graphicx}

\usepackage{url}

\newcommand\lhood{{\cal L}}

\newcommand\ptrue{f_{\rm true}}
\newcommand\ptruezero{f_{\rm true,0}}
\newcommand\ptrueone{f_{\rm true,1}}
\newcommand\hsat{H_{\rm sat}}
\newcommand\chip{\chi^2_{\rm P}}
\newcommand\chin{\chi^2_{\rm N}}

\newcommand\hatvecnu{\hat{\vec\nu}}
\newcommand\hatvecmu{\hat{\vec\mu}}
\newcommand\covmu{U}
\newcommand\hatcovmu{\hat U}
\newcommand\chicorr{\chi^2_{\rm corr}}

\begin{document}

\title{Should unfolded histograms be used to test hypotheses?}
\author{Robert D. Cousins\thanks{cousins@physics.ucla.edu} , 
Samuel J. May, and Yipeng Sun\\
Dept.\ of Physics and Astronomy\\ 
University of California, Los Angeles\\
Los Angeles, California 90095}
 
\date{July 24, 2016}

\maketitle

\begin{abstract}
In many analyses in high energy physics, attempts are made to remove
the effects of detector smearing in data by techniques referred to as
``unfolding'' histograms, thus obtaining estimates of the true values
of histogram bin contents.  Such unfolded histograms are then compared
to theoretical predictions, either to judge the goodness of fit of a
theory, or to compare the abilities of two or more theories to
describe the data.  When doing this, even informally, one is testing
hypotheses.  However, a more fundamentally sound way to test
hypotheses is to smear the theoretical predictions by simulating
detector response and then comparing to the data without unfolding;
this is also frequently done in high energy physics, particularly in
searches for new physics.  One can thus ask: to what extent does
hypothesis testing after unfolding data materially reproduce the
results obtained from testing by smearing theoretical predictions?  We
argue that this ``bottom-line-test'' of unfolding methods should be
studied more commonly, in addition to common practices of examining
variance and bias of estimates of the true contents of histogram bins.
We illustrate bottom-line-tests in a simple toy problem with two
hypotheses.
\end{abstract}

\section{Introduction}
\label{intro}

In high energy physics (HEP), unfolding (also called unsmearing) is a
general term describing methods that attempt to take out the effect of
smearing resolution in order to obtain a measurement of the true
underlying distribution of a quantity.  Typically the acquired data
(distorted by detector response, inefficiency, etc.) are binned in a
histogram.  The result of some unfolding procedure is then a new
histogram with estimates of the true mean bin contents prior to
smearing and inefficiency, along with some associated uncertainties.
It is commonly assumed that such unfolded distributions are useful
scientifically for comparing data to one or more theoretical
predictions, or even as quantitative measurements to be propagated
into further calculations.  Since an important aspect of the
scientific enterprise is to test hypotheses, we can ask: ``Should
unfolded histograms be used to test hypotheses?''  If the answer is
yes, then one can further ask if there are limitations to the utility
of testing hypotheses using unfolded histograms.  If the answer is no,
then the rationale for unfolding would seem to be limited.

In this note we illustrate an approach to answering the title question
with a few variations on a toy example that captures some of the
features of real-life unfolding problems in HEP.  The goal of the note
is to stimulate more interest in exploring what one of us (RC) has
called a {\em bottom-line test} for an unfolding method: {\em If the
  unfolded spectrum and supplied uncertainties are to be useful for
  evaluating which of two models is favored by the data (and by how
  much), then the answer should be materially the same as that which
  is obtained by smearing the two models and comparing directly to
  data without unfolding}~\cite{lyonsphystat2011}.  This is a
different emphasis for evaluating unfolding methods than that taken in
studies that focus on intermediate quantities such as bias and
variance of the estimates of the true mean contents, and on
frequentist coverage of the associated confidence intervals.  While
the focus here is on comparing two models for definiteness, the basic
idea of course applies to comparing one model to data (i.e., goodness
of fit), and to more general hypothesis tests.  Recently
Zech~\cite{zechtome} has extended the notion of the bottom-line test
to parameter estimation from fits to unfolded data, and revealed
failures in the cases studied, notably in fits to the width of a peak.

We adopt the notation of the monograph {\em Statistical Data Analysis}
by Glen Cowan \cite{cowan} (suppressing for simplicity the background
contribution that he calls $\vec\beta$):
\begin{description}
\item $y$ is a continuous variable representing the {\em true} value
  of some quantity of physical interest (for example momentum). It is
  distributed according to the pdf $\ptrue(y)$.
\item $x$ is a continuous variable representing the {\em observed}
  value of the same quantity of physical interest, after detector
  smearing effects and loss of events (if any) due to inefficiencies.
\item $s(x|y)$ is the resolution function of the detector: the
  conditional pdf for observing $x$, given that the true value is $y$
  (and given that it was observed somewhere).
\item $\vec \mu = (\mu_1, \dots, \mu_M)$ contains the expectation
  values of the bin contents of the {\em true} (unsmeared) histogram
  of $y$;
\item $\vec n = (n_1, \dots, n_N)$ contains the bin contents of the
  {\em observed} histogram (referred to as the {\em smeared
    histogram}, or occasionally as the {\em folded} histogram) of $x$
  in a single experiment;
\item $\vec \nu = (\nu_1, \dots, \nu_N)$ contains the expectation
  values of the bin contents of the {\em observed} (smeared) histogram
  of $x$, including the effect of inefficiencies: $\vec \nu = E[\vec
    n] $;
\item $R$ is the response matrix that gives the probability of an
  event in true bin $j$ being observed in bin $i$ after smearing:
  $R_{ij} = P({\rm observed~in~bin~}i | {\rm true~value~in~bin~}j)$;
\item $\hatvecmu = (\hat\mu_1, \dots, \hat\mu_M)$ contains the point
  estimates of $\vec\mu$ that are the output of an unfolding
  algorithm.
\item $\covmu$ is the covariance matrix of the estimates $\hatvecmu$:
  $U_{ij} = {\rm cov}[\hat\mu_i,\hat\mu_j]$.  The estimate of $\covmu$
  provided by an unfolding algorithm is $\hatcovmu$.

\end{description}
Thus we have  
\begin{equation}
\label{nuRmu}
\vec\nu = R\,\vec\mu.
\end{equation}
As discussed by Cowan and noted above, $R$ includes the effect of the
efficiency $\epsilon$, i.e., the effect of events in the true
histograms not being observed in the smeared histogram. The only
efficiency effect that we consider here is that due to events being
smeared outside the boundaries of the histogram.  (That is, we do not
consider an underflow bin or an overflow bin.)

The response matrix $R$ depends on the resolution function and on
(unknown) true bin contents (and in particular on their true densities
$\ptrue(y)$ {\em within} each bin), and hence $R$ is either known only
approximately or as a function of assumptions about the true bin
contents.  The numbers of bins $M$ and $N$ need not be the
same. ($N>M$ is often suggested, while $N<M$ leaves the system of
equations under-determined.)  For the toy studies discussed here, we
set $N=M=10$, so that $R$ is a square matrix that typically has an
inverse.

In the smeared space, we take the observed counts $\vec n$ to be
independent observations from the underlying Poisson distributions:
\begin{equation}
\label{poisprob}
P(n_i;\nu_i) = \frac{\nu_i^{n_i}\exp(-\nu_i)}{n_i!}.
\end{equation}
The unfolding problem is then to use $R$ and $\vec n$ as inputs to
obtain estimates $\hatvecmu$ of $\vec\mu$, and to obtain the
covariance matrix $\covmu$ of these estimates (or rather an estimate
of $\covmu$, $\hatcovmu$), ideally taking in account uncertainty in
$R$.

When reporting unfolded results, authors report $\hatvecmu$, ideally
along with $\hatcovmu$. (If only a histogram of $\hatvecmu$ with
``error bars'' is displayed, then only the diagonal elements of
$\hatcovmu$ are communicated, further degrading the information.)  The
``bottom line test'' of an application of unfolding is then whether
hypothesis tests about underlying models that predict $\vec\mu$ can
obtain meaningful results if they take as input $\hatvecmu$ and
$\hatcovmu$.

\section{The toy models}
\begin{figure}
\begin{center}
\includegraphics[width=0.49\textwidth]{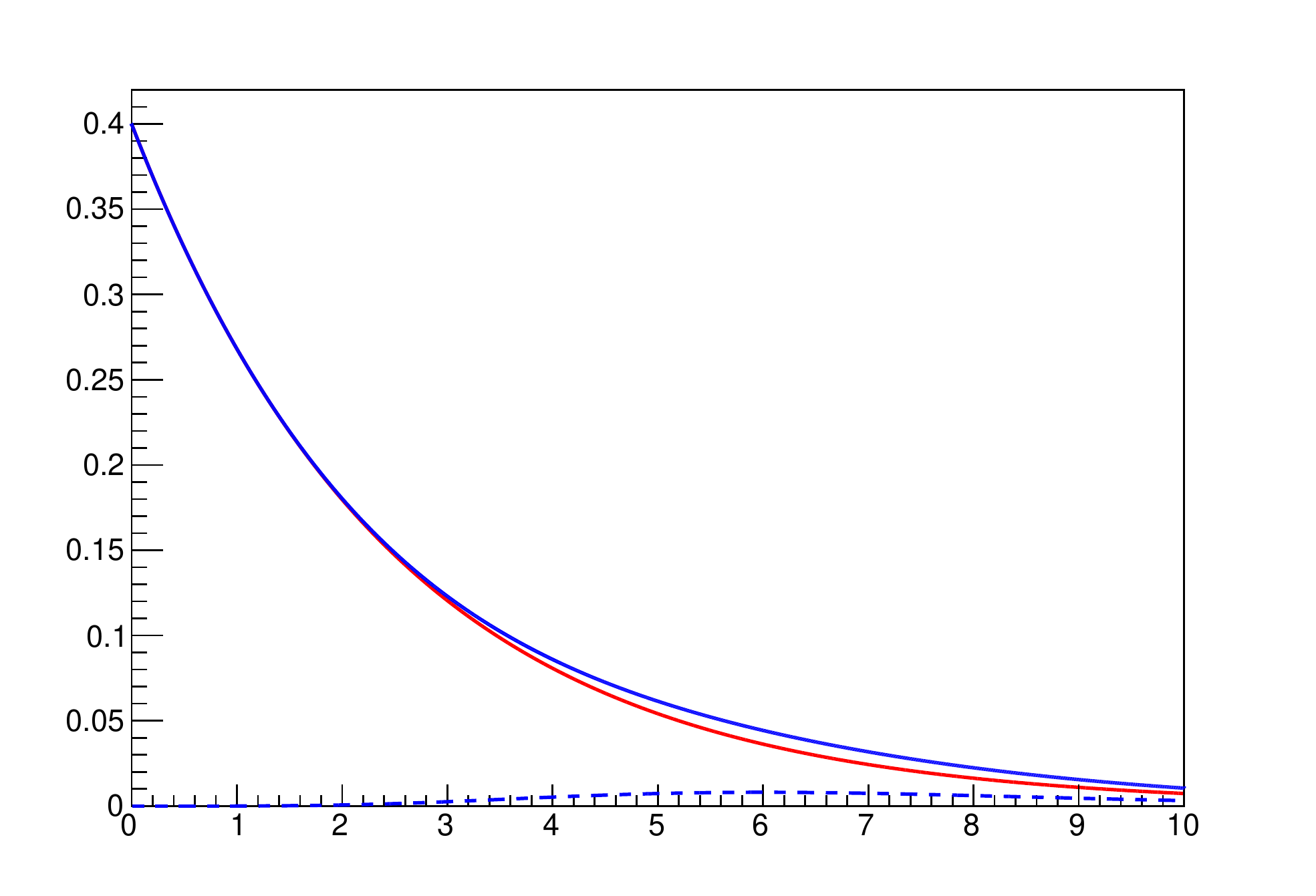}
\includegraphics[width=0.49\textwidth]{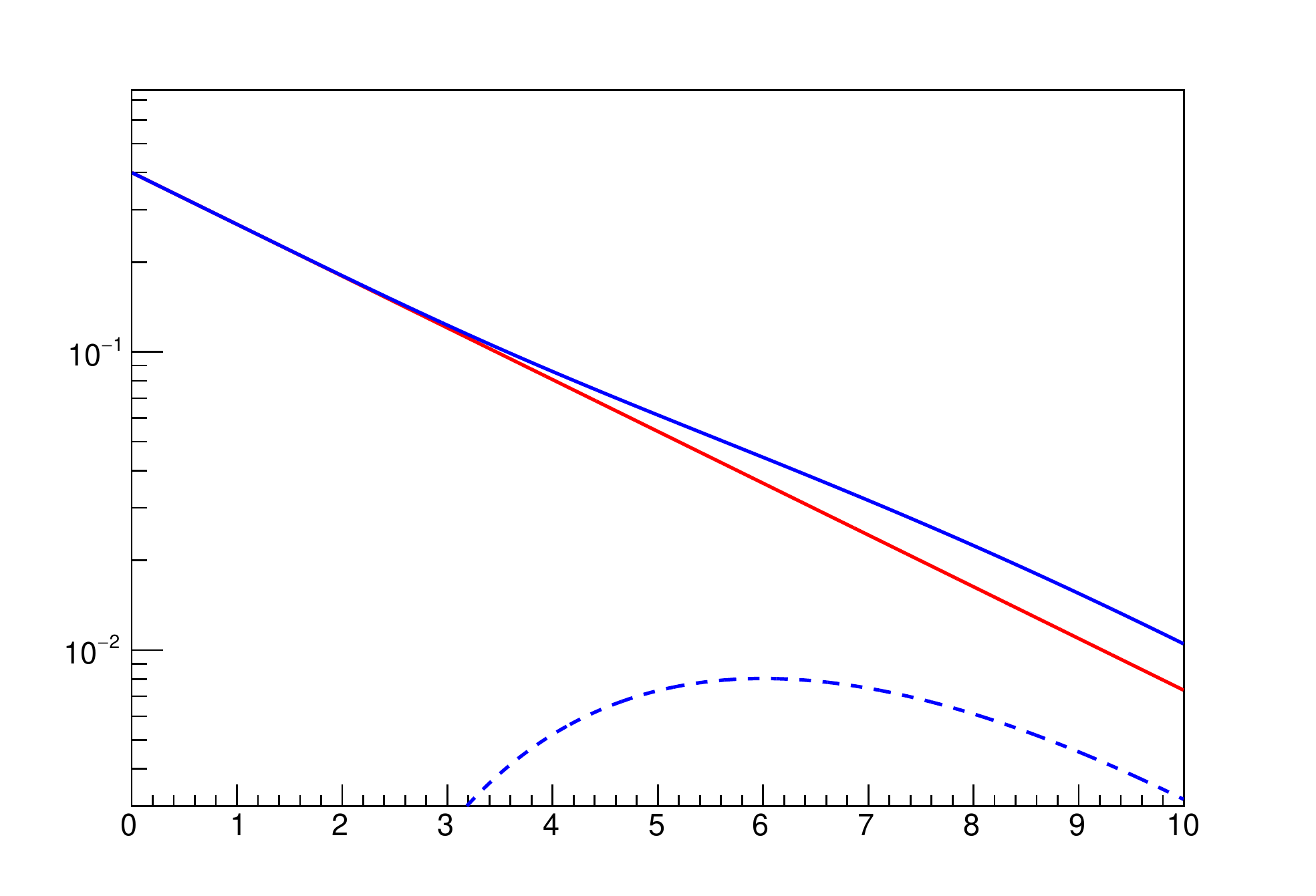}
\caption{Baseline pdfs in the study, with (left) linear vertical scale
  and (right) logarithmic vertical scale.  The null hypothesis $H_0$
  is represented by $\ptruezero(y)$, shown in red.  The alternative
  hypothesis $H_1$ has an additional component shown in dashed blue,
  with the sum $\ptrueone(y)$ in solid blue.
%-------truthsLin.pdf truthsLog
}
\label{truepdfs}
\end{center}
\end{figure}

For the null hypothesis $H_0$, we consider the continuous variable $y$
to be distributed according the true pdf
\begin{equation}
\label{nullp}
\ptruezero(y)  = A \exp(-y/\tau),
\end{equation} 
where $\tau$ is known, and $A$ is a normalization constant.  For the
alternative hypothesis $H_1$, we consider $y$ to be distributed
according the true pdf
\begin{equation}
\label{altp}
\ptrueone(y)  = A \left( \exp(-y/\tau) + Bg(y) \right),
\end{equation}
where $\tau$ is the same as in the null hypothesis, and where $g(y)$
is a pdf that encodes a departure from the null hypothesis.  In this
note, we assume that both $g(y)$ and $B$ are known, and lead to
potentially significant departures from the null hypothesis at large
$y$.  The constant $B$ controls the level of such departures.
Figure~\ref{truepdfs} displays the baseline pdfs that form the basis
of the current study, for which we take $g$ to be a normalized gamma
distribution,
\begin{equation}
\label{gdef}
g(y) = y^6\,\exp(-y)/6!
\end{equation}
and $B=0.05$.

For each hypothesis, the true bin contents $\vec\mu$ are then each
proportional to the integral of the relevant $\ptrue(y)$ over each
bin.  For both hypotheses, we take the smearing of $x$ to be the
Gaussian resolution function,
\begin{equation}
\label{smear}
s(x|y) =    \frac{1}{\sqrt{2\pi}\sigma}\exp(-(x-y)^2/2\sigma^2),
\end{equation}
where $\sigma$ is known.  

\begin{figure}
\begin{center}
\includegraphics[width=0.49\textwidth]{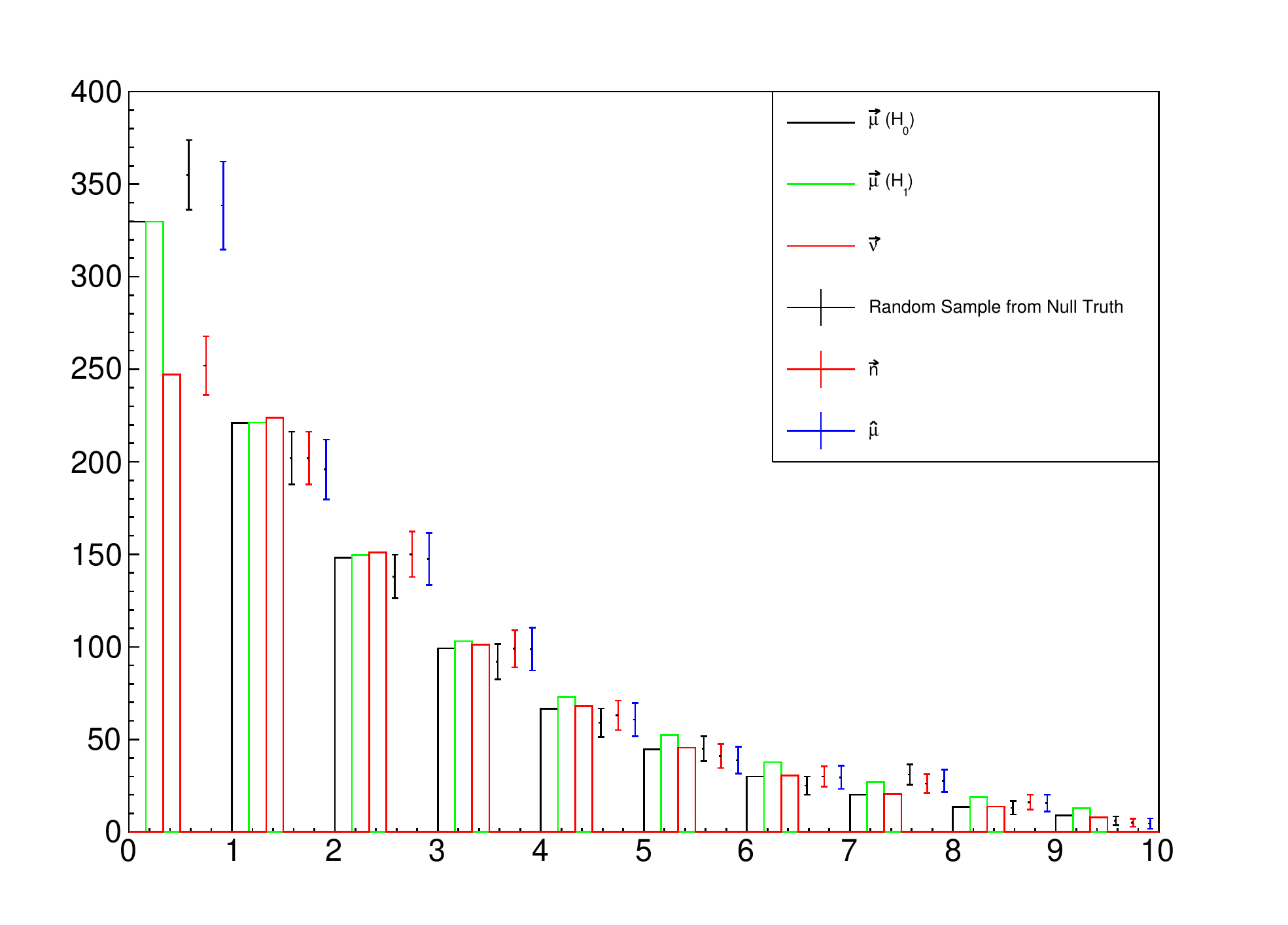}
\includegraphics[width=0.49\textwidth]{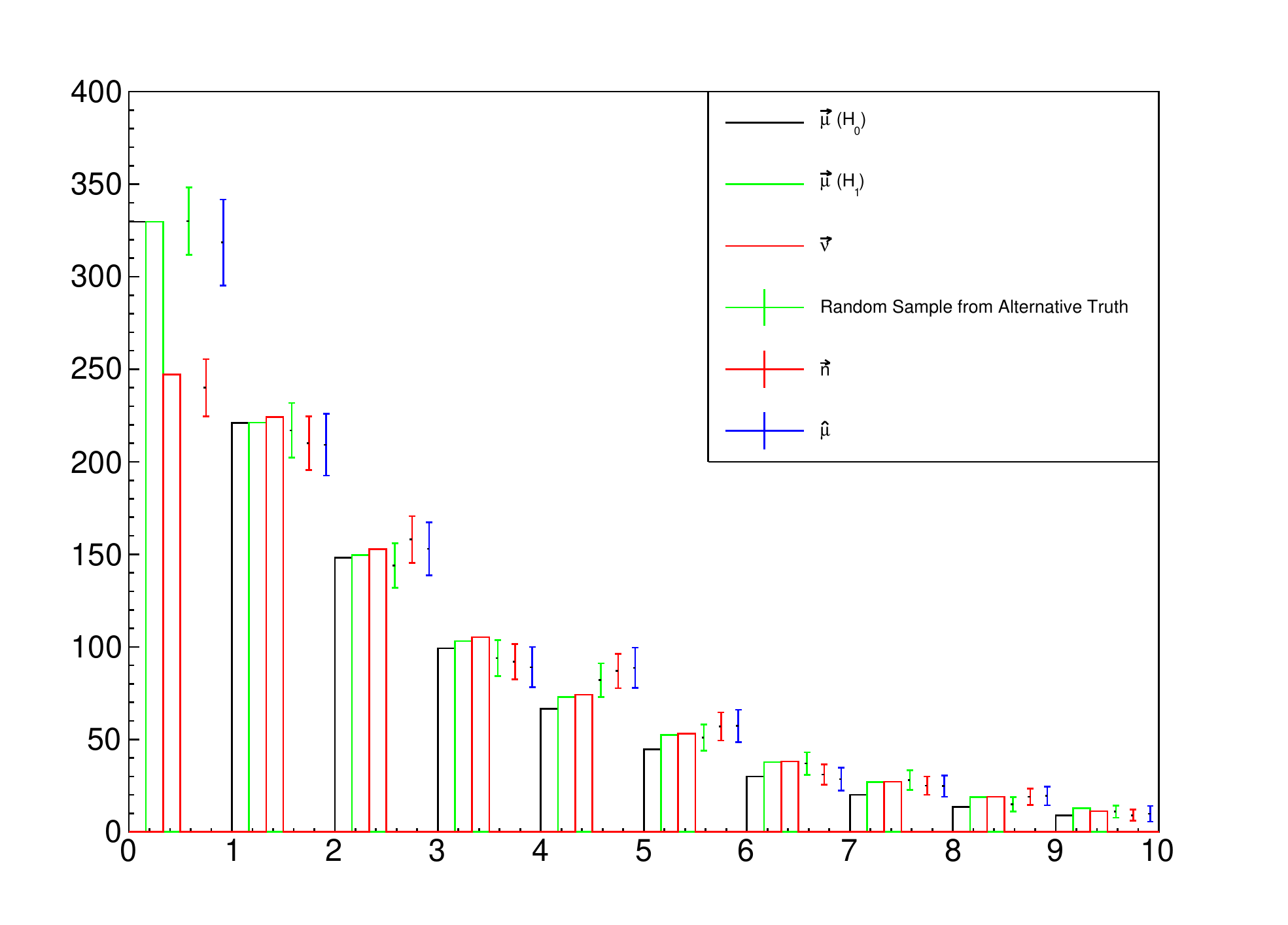}
\caption{Solid histograms: the true bin contents for unsmeared
  $\vec\mu$ and smeared $\vec\nu$, for (left) the null hypothesis
  $H_0$ and (right) the alternative hypothesis $H_1$.  Data points: In
  MC simulation a set $\{y\}$ of true points is chosen randomly and
  then smeared to be the set $\{x\}$.  The three points plotted in
  each bin are then the bin contents when $y$ and $x$ are binned,
  followed by the unfolded estimate for bin contents.
%----------UnfoldBasicIterative.pdf Unfold\_TwoHypothesis\_TwoExp.pdf
}
\label{histos}
\end{center}
\end{figure}

\begin{figure}
\begin{center}
\includegraphics[width=0.49\textwidth]{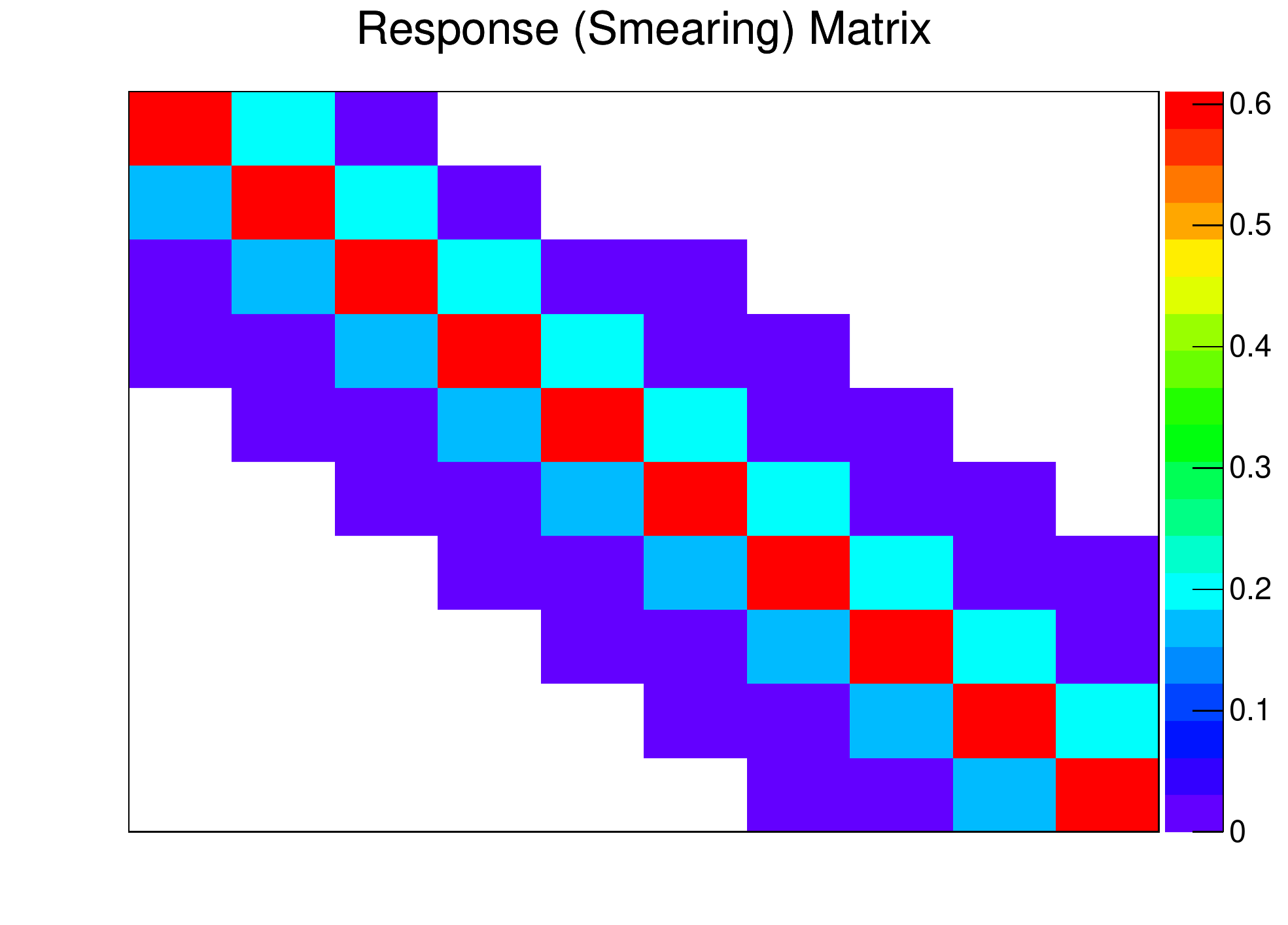}
\includegraphics[width=0.49\textwidth]{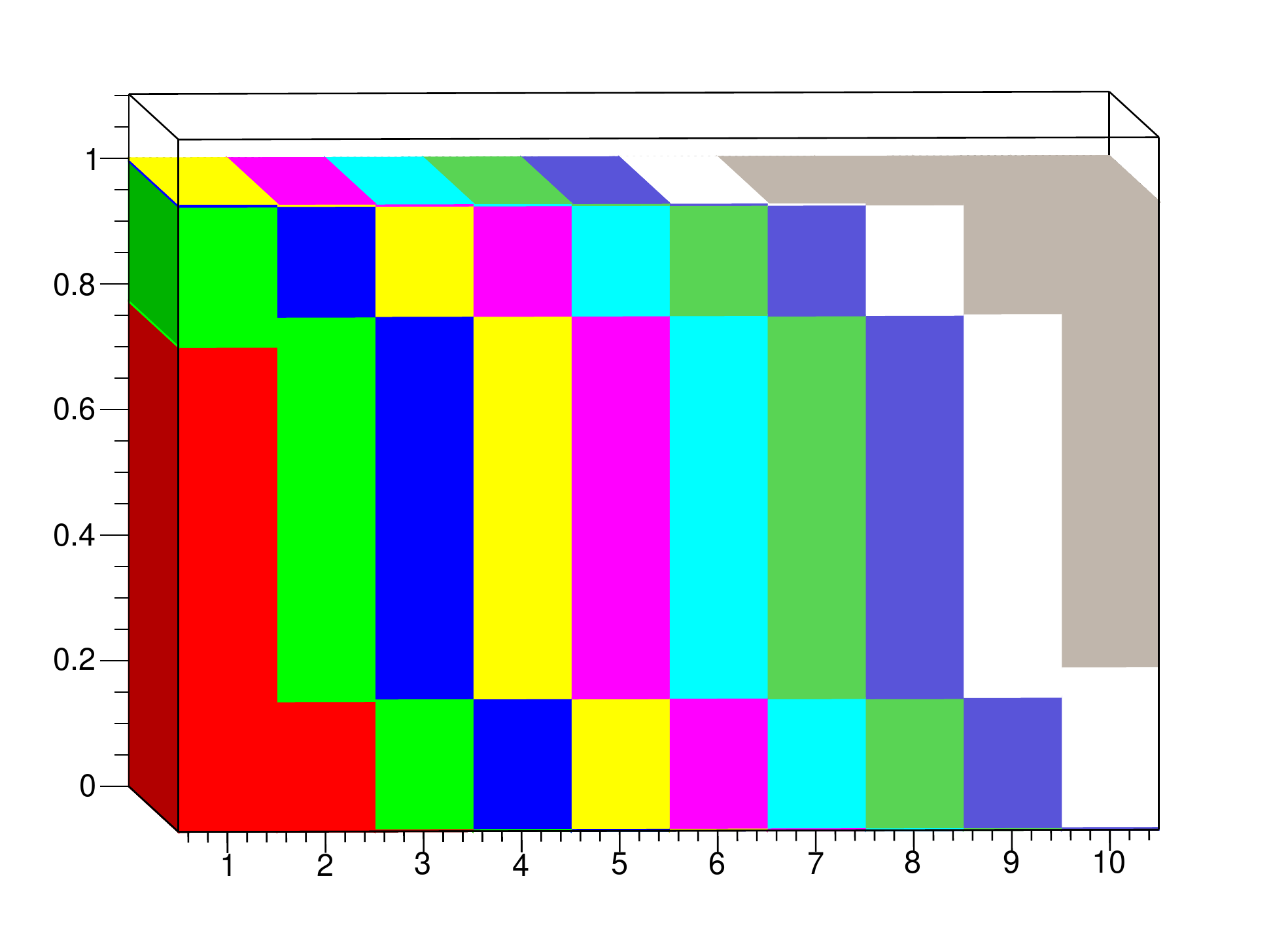}
\caption{(left) The response matrix $R$ for default parameter values
  in Table~\ref{baseline}. (right) For each bin in the measured $y$
  value, the fraction of events that come from that bin (dominant
  color) and from nearby bins.
%----------- Matrices\_BinPurity 3 4
}
\label{responsepurity}
\end{center}
\end{figure}

For baseline plots, we use the values shown in Table~\ref{baseline},
and the study the effect of varying one parameter at a time.  For both
$x$ and $y$, we consider histograms with 10 bins of width 1 spanning
the interval [0,10].  The default $\sigma$ is half this bin width.
The quantities $\vec\mu$, $R$, and $\vec\nu$ are then readily computed
as in Ref.~\cite{cowan}.  Figure~\ref{histos} displays $\vec\mu$ and
$\vec\nu$ (in solid histograms), while Fig.~\ref{responsepurity}
displays the response matrix as well as the source bin of events that
are observed in each bin.  In each simulated experiment, the total
number of events is sampled from a Poisson distribution with mean
given in Table~\ref{baseline}.

\begin{table}[h]
\begin{center}
\caption{Values of parameters used in the baseline unfolding examples
}
\label{baseline}
\begin{tabular}{lll} \hline
Parameter &Baseline value \\ \hline
Amplitude $B$ of departure from null & 0.05 \\
Exponential decay parameter $\tau$ & 2.5 \\
Mean number of events in each simulated experiment &  1000    \\ 
Number of histogram bins & 10 \\
Bin width & 1.0 \\
Gaussian $\sigma$ for smearing &  0.5  \\
Number of events used to construct $R$ &  $10^7$  \\
Number of iterations in EM unfolding &  4	  \\\hline
\end{tabular}
\end{center}
\end{table}

Boundary effects at the ends of the histogram are an important part of
a real problem.  In our simplified toy problems, we use the same
smearing for events near boundaries as for all events (hence not
modeling correctly some physical situations where observed values
cannot be less than zero); events that are smeared to values outside
the histogram are considered lost and contribute to the inefficiencies
included in $R$.

These toy models capture some important aspects of real problems in
HEP.  For example, one might be comparing event generators for
top-quark production in the Standard Model.  The variable $y$ might be
the transverse momentum of the top quark, and the two hypotheses might
be two calculations, one to higher order.

Another real problem might be where $y$ represents transverse momentum
of jets, the null hypothesis is the standard model, and the
alternative hypothesis is some non-standard-model physics that turns
on at high transverse momentum. (In this case, it is typically not the
case that amplitude $B$ of additional physics is known.)

\section{Hypothesis tests in the smeared space}

In a typical search for non-standard-model physics, the hypothesis
test of $H_0$ vs.\ $H_1$ is formulated in the smeared space, i.e., by
comparing the histogram contents $\vec n$ to the mean bin contents
$\vec\nu$ predicted by the true densities $\ptrue(y)$ under each
hypothesis combined with the resolution function and any efficiency
losses.  The likelihood $\lhood(H_0)$ for the null hypothesis is the
product over bins of the Poisson probability of obtaining the observed
bins counts:
\begin{equation}
\label{likhood}
 \lhood(H_0)  = \prod_i P(n_i;\nu_i),
\end{equation}
where the $\nu_i$ are taken from the null hypothesis
prediction. Likelihoods for other hypotheses, such as $\lhood(H_1)$,
are constructed similarly.

For testing goodness of fit, it can be useful~\cite{bakercousins,PDG}
to use the observed data to construct a third hypothesis, $\hsat$,
corresponding the {\em saturated model}~\cite{lindsey}, which sets the
predicted mean bin contents to be exactly those observed.  Thus
$\lhood(\hsat)$ is the upper bound on $\lhood(H)$ for any hypothesis,
given the observed data.  The negative log-likelihood ratio
\begin{equation}
\label{baker}
-2\ln \lambda_{0,{\rm sat}} = -2\ln\left(\frac{\lhood(H_0)}{\lhood(\hsat)}\right)
\end{equation} 
is a goodness-of-fit test statistic that is asymptotically distributed
as a chisquare distribution if $H_0$ is true. Similarly one has
$-2\ln\lambda_{1,{\rm sat}}$ for testing $H_1$.

An alternative (in fact older) goodness-of-fit test statistic is
Pearson's chisquare~\cite{PDG},
\begin{equation}
\label{pearson}
\chip = \sum_{i=1}^N \frac{(n_i - \nu_i)^2}{\nu_i}.
\end{equation}
Yet another alternative, generally less favored, is known as Neyman's
chisquare~\cite{bakercousins},
\begin{equation}
\label{neyman}
\chin = \sum_{i=1}^N \frac{(n_i - \nu_i)^2}{n_i}.
\end{equation}
Ref.~\cite{bakercousins} argues that Eqn.~\ref{baker} is the most
appropriate GOF statistic for Poisson-distributed histograms, and we
use it as our reference point in the smeared space.

\begin{figure}
\begin{center}
\includegraphics[width=0.49\textwidth]{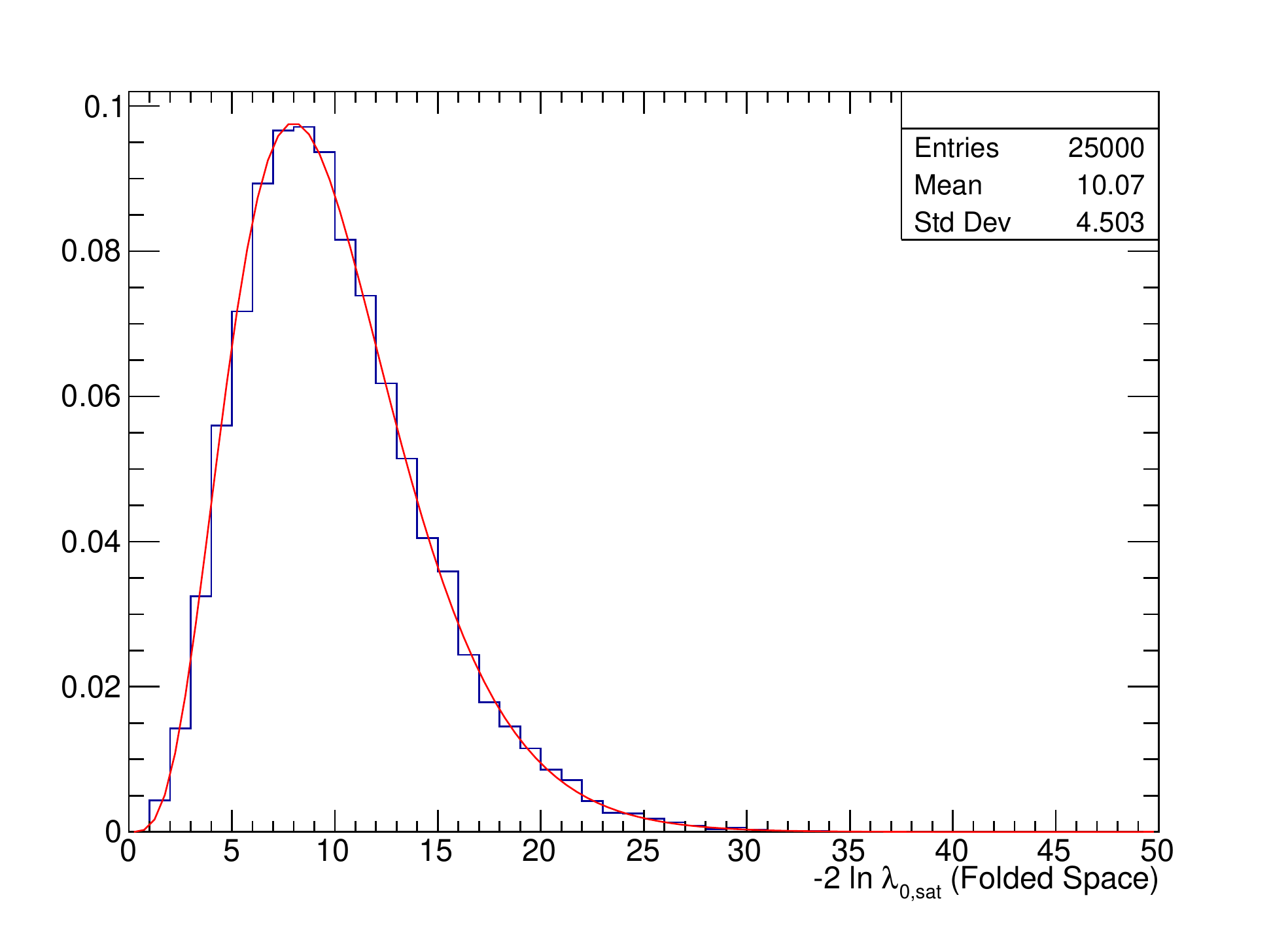}
\includegraphics[width=0.49\textwidth]{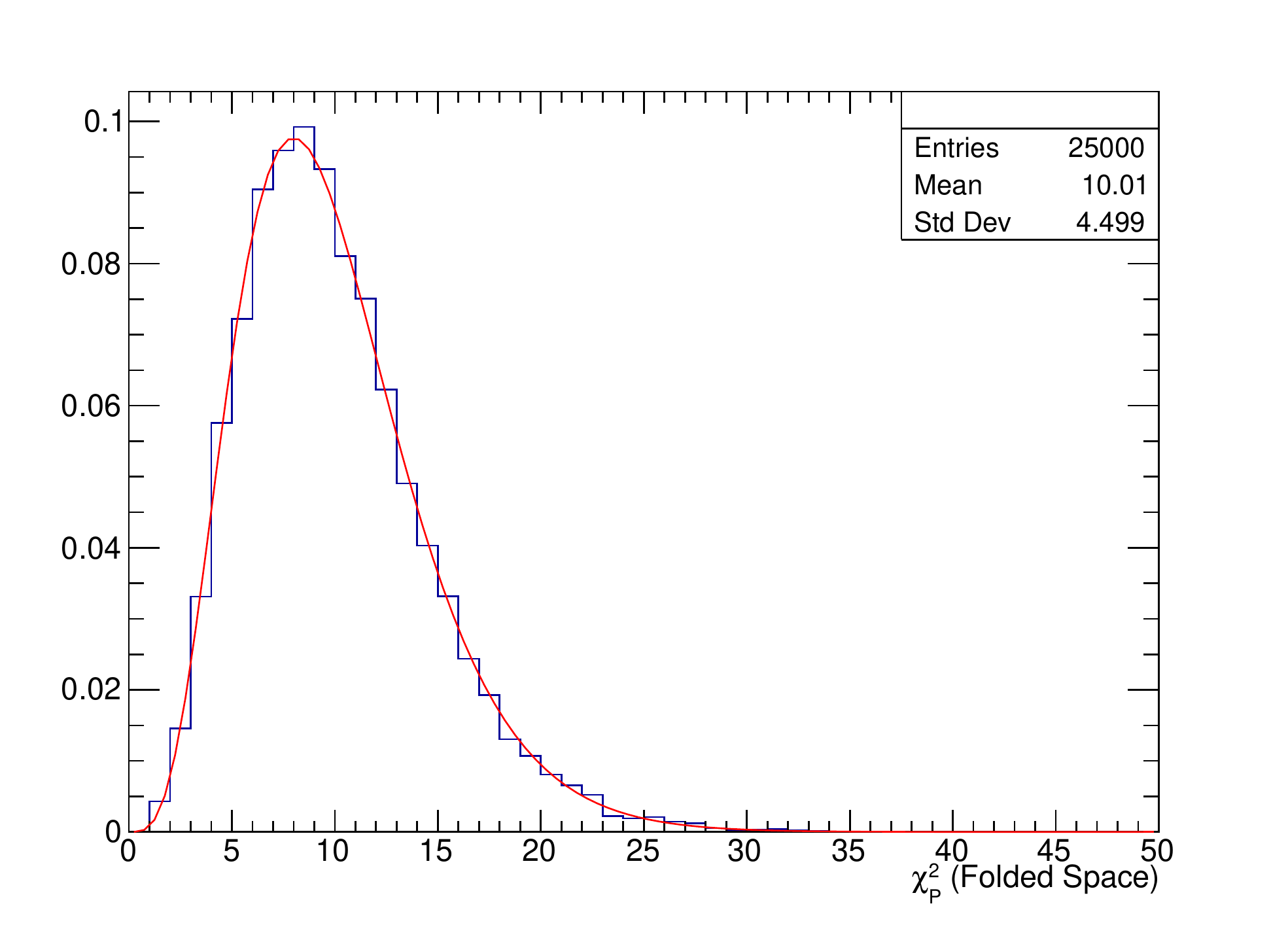}
\includegraphics[width=0.49\textwidth]{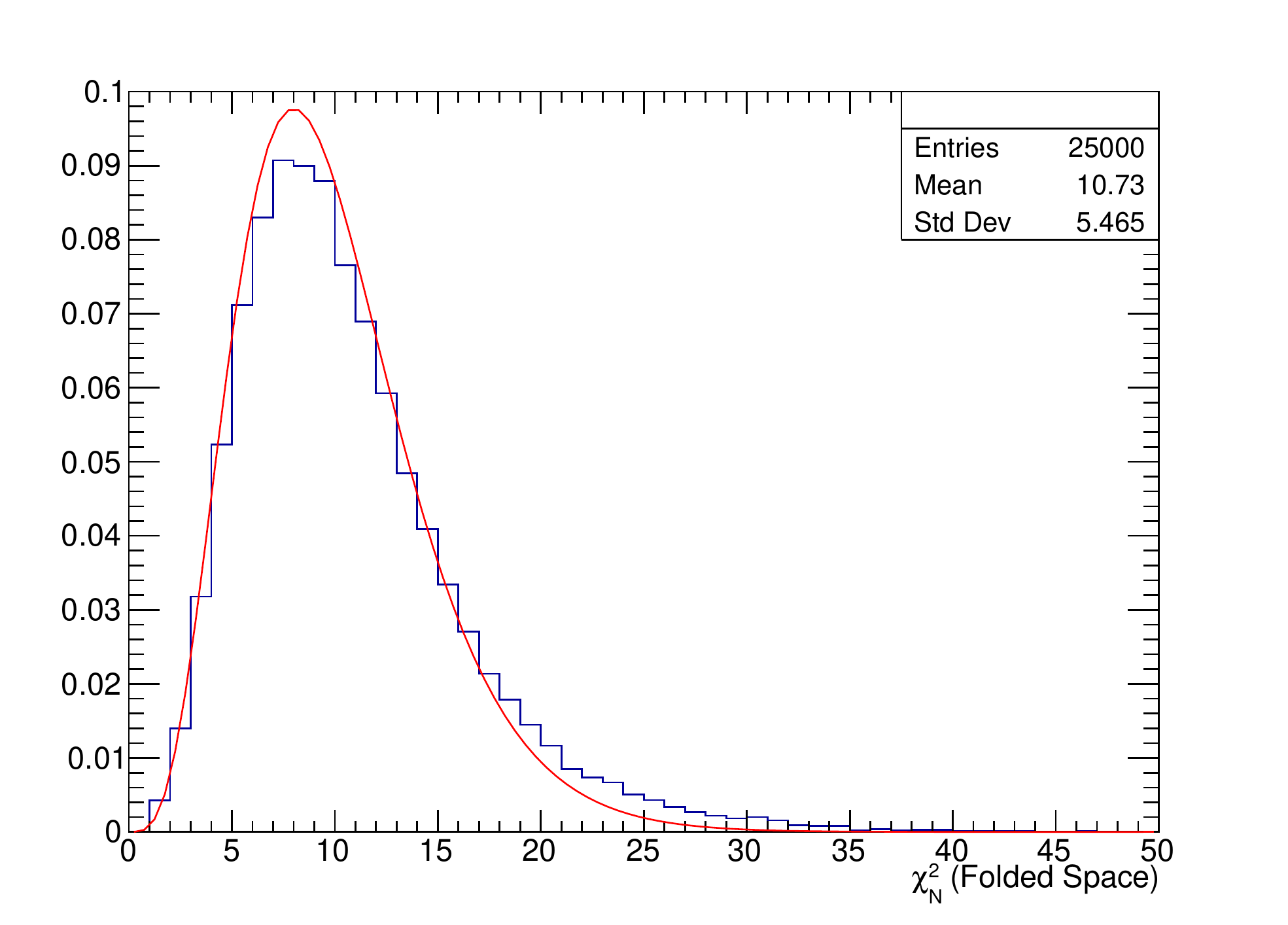}
\includegraphics[width=0.49\textwidth]{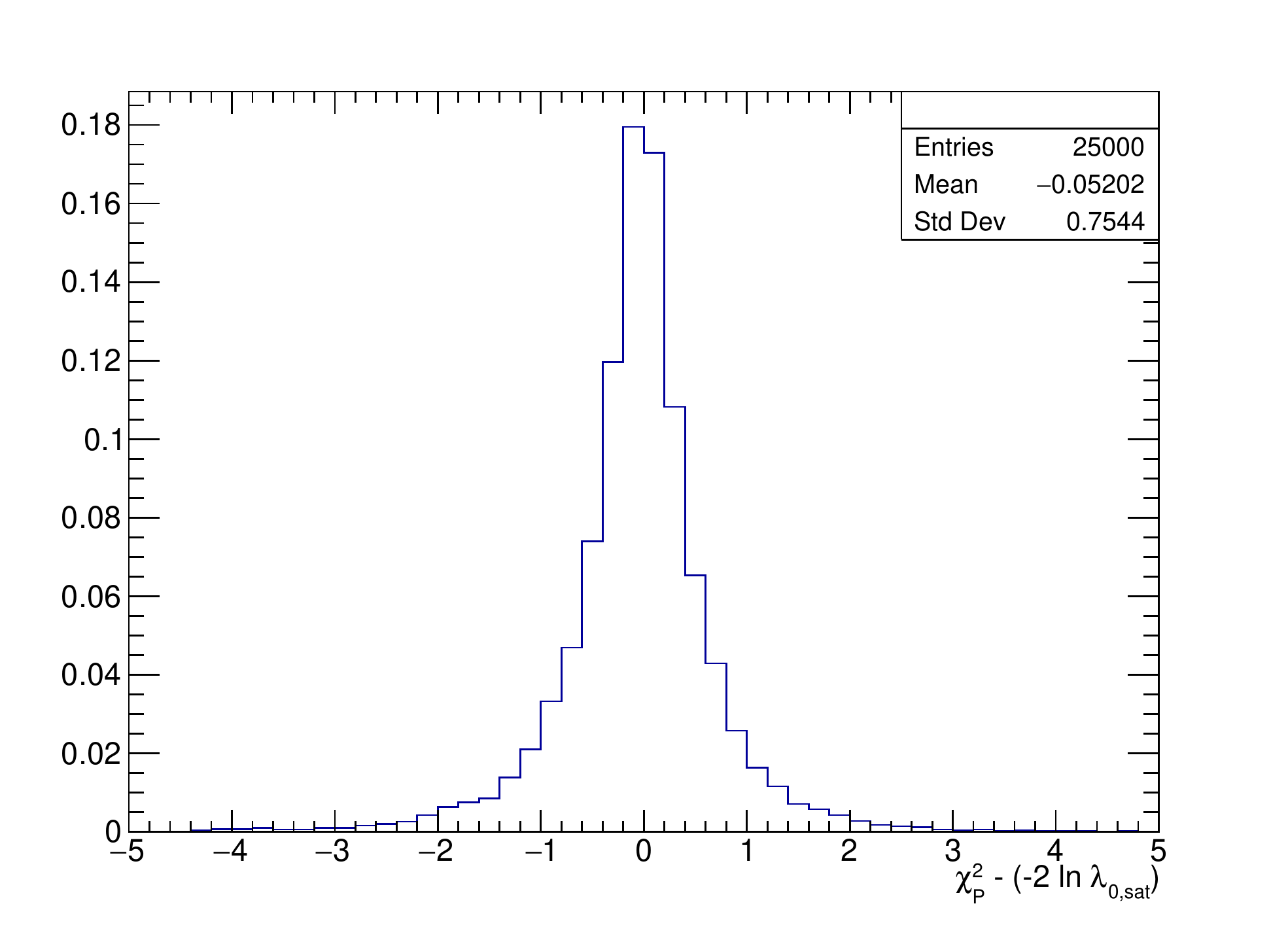}
\caption{For events generated under $H_0$, in the smeared space with
  default value of Gaussian $\sigma$, histograms of the GOF test
  statistics: (top left) $-2\ln\lambda_{0,{\rm sat}}$, (top right)
  $\chip$, and (bottom left) $\chin$.  The solid curves are the
  chisquare distribution with 10 DOF.  (bottom right) Histogram of the
  event-by-event difference in the two GOF test statistics $\chip$ and
  $-2\ln\lambda_{0,{\rm sat}}$.
%--------- OneHypothesis\_1DSlice 1 2 4
%--------- OneTruth\_2DChi2Histograms.pdf 1 
}
\label{nullgofsmeared}
\end{center}
\end{figure}

Figure~\ref{nullgofsmeared} shows the distributions of
$-2\ln\lambda_{0,{\rm sat}}$ and $\chip$, and their difference, for
histograms generated under $H_0$. Both distributions follow the
expected $\chi^2$ distribution with 10 degrees of freedom (DOF).  In
contrast, the histogram of $\chin$ (Figure~\ref{nullgofsmeared}
(bottom left)) has noticeable differences from the theoretical curve.

\begin{figure}
\begin{center}
\includegraphics[width=0.49\textwidth]{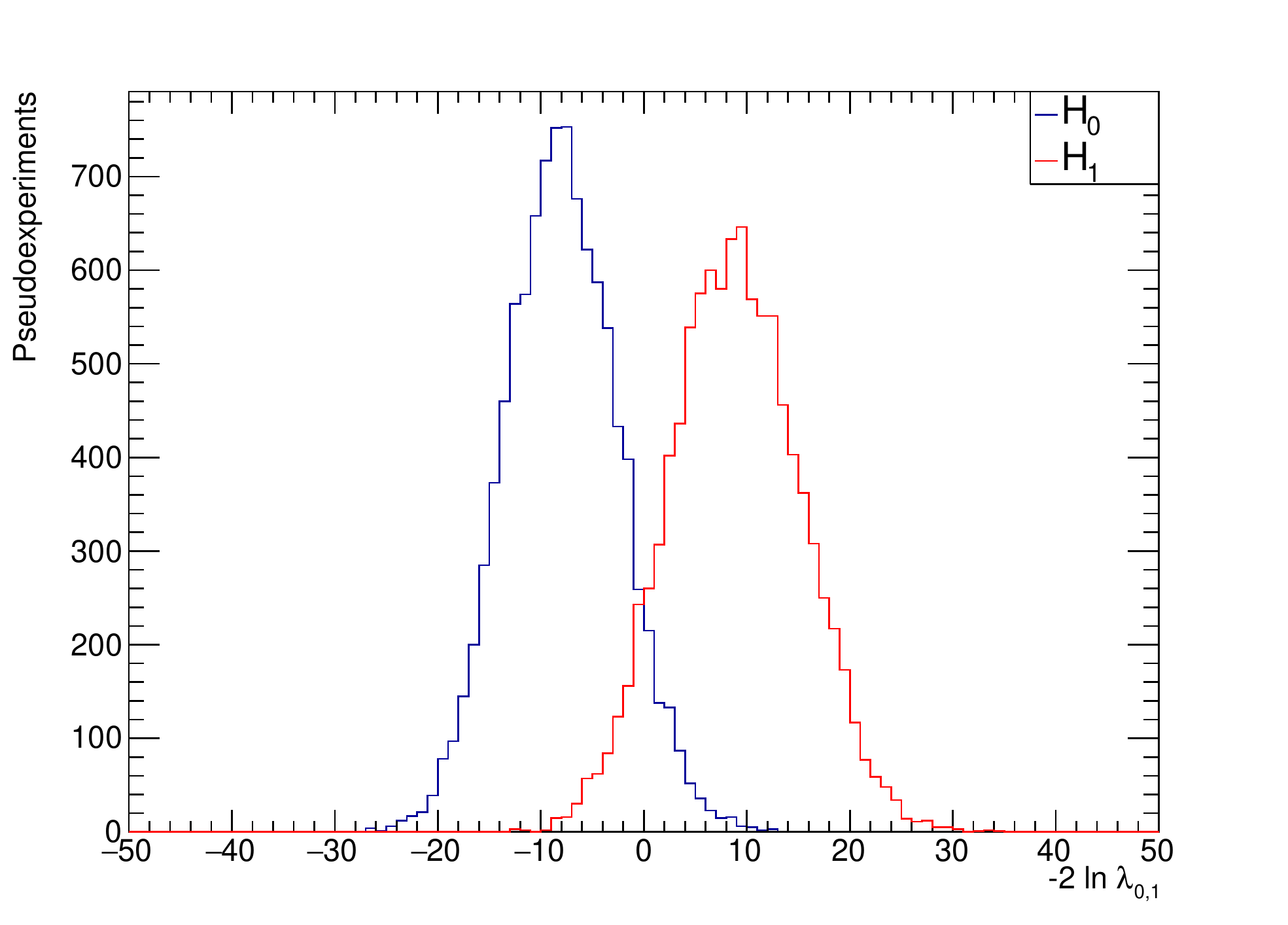}
\caption{In the smeared space, histogram of the test statistic 
$-2\ln\lambda_{0,1}$ for
events generated under $H_0$ (in blue) and $H_1$ (in red). 
%--------- HypothesisComparison 1
}
\label{lambdah0h1}
\end{center}
\end{figure}

For testing $H_0$ vs.\ $H_1$, a suitable test statistic is the
likelihood ratio $\lambda$ formed from the probabilities of obtaining
bin contents $\vec n$ under each hypothesis:
\begin{equation}
\label{lambda}
-2\ln \lambda_{0,1} = -2\ln\left(\frac{\lhood(H_0)}{\lhood(H_1)}\right)
 = -2\ln\lambda_{0,{\rm sat}}  - (-2\ln\lambda_{1,{\rm sat}}\,),
\end{equation} 
where the second equality follows from Eqn.~\ref{baker}.
Figure~\ref{lambdah0h1} shows the distribution of $-2\ln\lambda_{0,1}$
for events generated under $H_0$ and for events generated under $H_1$,
using the default parameter values in Table~\ref{baseline}.

We would assert that these results obtained in the smeared space are
the ``right answers'' for chisquare-like GOF tests of $H_0$ and $H_1$
(if desired), and in particular for the likelihood-ratio test of $H_0$
vs $H_1$ in Fig.~\ref{lambdah0h1}.  Given a particular observed data
set, such histograms can be used to calculate $p$-values for each
hypothesis, simply by integrating the appropriate tail of the
histogram beyond the observed value of the relevant likelihood
ratio~\cite{PDG}. In frequentist statistics, such $p$-values are
typically the basis for inference, especially for the simple-vs-simple
hypothesis tests considered here. (Of course there is a vast
literature questioning the foundations of using $p$-values, but in
this note we assume that they can be useful, and are interested in
comparing ways to compute them.)

We compare $-2\ln\lambda$, $\chip$, $\chin$, and the generalization of
Eqn.~\ref{chisq} including correlations in various contexts below.
For Poisson-distributed data, arguments in favor of $-2\ln\lambda$
when it is available are in Ref.~\cite{bakercousins}.

\subsection{Note regarding Gaussian data and the usual $\chi^2$ GOF test}
In the usual $\chi^2$ GOF test with (uncorrelated) estimates
$\hatvecnu$ having {\em Gaussian} densities with standard deviations
$\vec\sigma$, one would commonly have
\begin{equation}
\label{chisq}
\chi^2 = \sum_{i=1}^N \frac{(\hat\nu_i - \nu_i)^2}{\sigma_i^2}.
\end{equation}
Although not usually mentioned, this is equivalent to a likelihood
ratio test with respect to the saturated model, just as in the Poisson
case.  The likelihood is
\begin{equation}
 \lhood  =  \prod_i \frac{1}{\sqrt{2\pi\sigma_i^2}}
            \exp\left(-(\hat\nu_i - \nu_i)^2/2\sigma_i^2\right),
 \label{like}
\end{equation}
where for $\lhood(H_0)$ one has $\nu_i$ predicted by $H_0$,
and for the saturated model, one has $\nu_i  = \hat\nu_i$.  Thus 
\begin{equation}
 \lhood(H_{\rm sat})  =  \prod_i \frac{1}{\sqrt{2\pi\sigma_i^2}},
\end{equation}
and hence $\chi^2 = -2\ln(\lhood(H_0)/\lhood(H_{\rm sat})) =
-2\ln\lambda_{0,{\rm sat}}.$ (It is sometimes said loosely and
incorrectly that for the Gaussian model, $\chi^2 = -2\ln\lhood(H_0)$,
but clearly the ratio is necessary to cancel the normalization
factor.)

There is also a well-known connection between the usual Gaussian
$\chi^2$ of Eqn.~\ref{chisq} and Pearson's chisquare in
Eqn.~\ref{pearson}: since the variance of a Poisson distribution is
equal to its mean, a naive derivation of Eqn.~\ref{pearson} follows
immediately from Eqn.~\ref{chisq}.  If one further approximates
$\nu_i$ by the estimate $n_i$, then one obtains Neyman's chisquare in
Eqn.~\ref{neyman}.

\section{Unfolding by using maximum likelihood estimates and
approximations from truncated iterative solutions}

If one unfolds histograms and then compares the unfolded histograms
$\hatvecmu$ to (never smeared) model predictions $\mu$, even
informally, then one is implicitly assuming that the comparison is
scientifically meaningful.  For this to be the case, we would assert
that the results of comparisons should not differ materially from the
``right answers'' obtained above in the smeared space. Here we explore
a few test cases.

Given the observed histogram contents $\vec n$, the likelihood
function for the unknown $\vec \nu$ follows from Eqn.~\ref{poisprob}
and leads to the maximum likelihood (ML) estimates $\hat\nu_i = n_i$,
i.e.,
\begin{equation}
\label{nun}
\hatvecnu = \vec n.
\end{equation}
One might then expect that the ML estimates of the unknown means
$\vec\mu$ can be obtained by substituting $\hatvecnu$ for $\vec n$ in
Eqn.~\ref{nuRmu}.  If $R$ is a square matrix, as assumed here, then
this yields
\begin{equation}
\label{nuRmuinv}
\hatvecmu = R^{-1}\,\hatvecnu = R^{-1}\,\vec n.
\end{equation}
These are indeed the ML estimates of $\vec\mu$ as long as $R$ is
invertible and the estimates $\mu_i$ are
positive~\cite{cowan,kuusela}, which is generally the case in the toy
problem studied here.

The covariance matrix of the estimates $\hatvecmu$ in terms of $R$ and
$\vec \nu$ is derived in Ref.~\cite{cowan}:
\begin{equation}
\label{covmu}
\covmu = R^{-1}\, V\, (R^{-1})^T,
\end{equation}
where $V_{ij} = \delta_{ij}\nu_i$.  Since the true values $\vec \nu$
are presumed unknown, it is natural to substitute the estimates from
Eqn.~\ref{nun}, thus obtaining an estimate $\hatcovmu$.  Consequences
of this approximation are discussed below.

In all cases (even when matrix inversion fails), the ML estimates for
$\vec\mu$ can be found to desired precision by the iterative method
variously known as \cite{kuusela} Expectation Maximization (EM),
Lucy-Richardson, or (in HEP) the iterative method of
D'Agostini~\cite{dagostini}.  Because the title of
Ref.~\cite{dagostini} mentions Bayes' Theorem, in HEP the EM method is
unfortunately (and wrongly) referred to as ``Bayesian'', even though
it is a fully frequentist algorithm~\cite{kuusela}.

As discussed by Cowan~\cite{cowan}, the ML estimates are unbiased, but
the unbiasedness can come at a price of large variance that renders
the unfolded histogram unintelligible to humans.  Therefore there is a
vast literature on ``regularization methods'' that reduce the variance
at the price of increased bias, such that the mean-squared-error (the
sum of the bias squared and the variance) is (one hopes) reduced.

The method of regularization popularized in HEP by
D'Agostini~\cite{dagostini} (and studied for example by Bohm and
Zech~\cite{bohmzech}) is simply to stop the iterative EM method before
it converges to the ML solution.  The estimates $\hat{\vec \mu}$ then
retain some memory of the starting point of the solution (typically
leading to a bias) and have lower variance.  The uncertainties
(covariance matrix) also depend on when the iteration stops.

Our studies in this note focus on the ML and truncated iterative EM
solutions, and use the EM implementation (unfortunately called
RooUnfoldBayes) in the RooUnfold~\cite{roounfold} suite of unfolding
tools.  This means that for the present studies, we are constrained by
the policy in RooUnfold to use the ``truth'' of the training sample to
be the starting point for the iterative EM method; thus we have not
studied convergence starting from, for example, a uniform
distribution.  Useful studies of the bias of estimates are thus not
performed.

Other popular methods in HEP include variants of Tikhonov
regularization, such as ``SVD'' method advocated by Hocker and
Kartvelishvili~\cite{hocker}, and the implementation included in
TUnfold~\cite{tunfold}.  The relationship of these methods to those in
the professional statistics literature is discussed by
Kuusela~\cite{kuusela}.

\begin{figure}
\begin{center}
\includegraphics[width=0.49\textwidth]{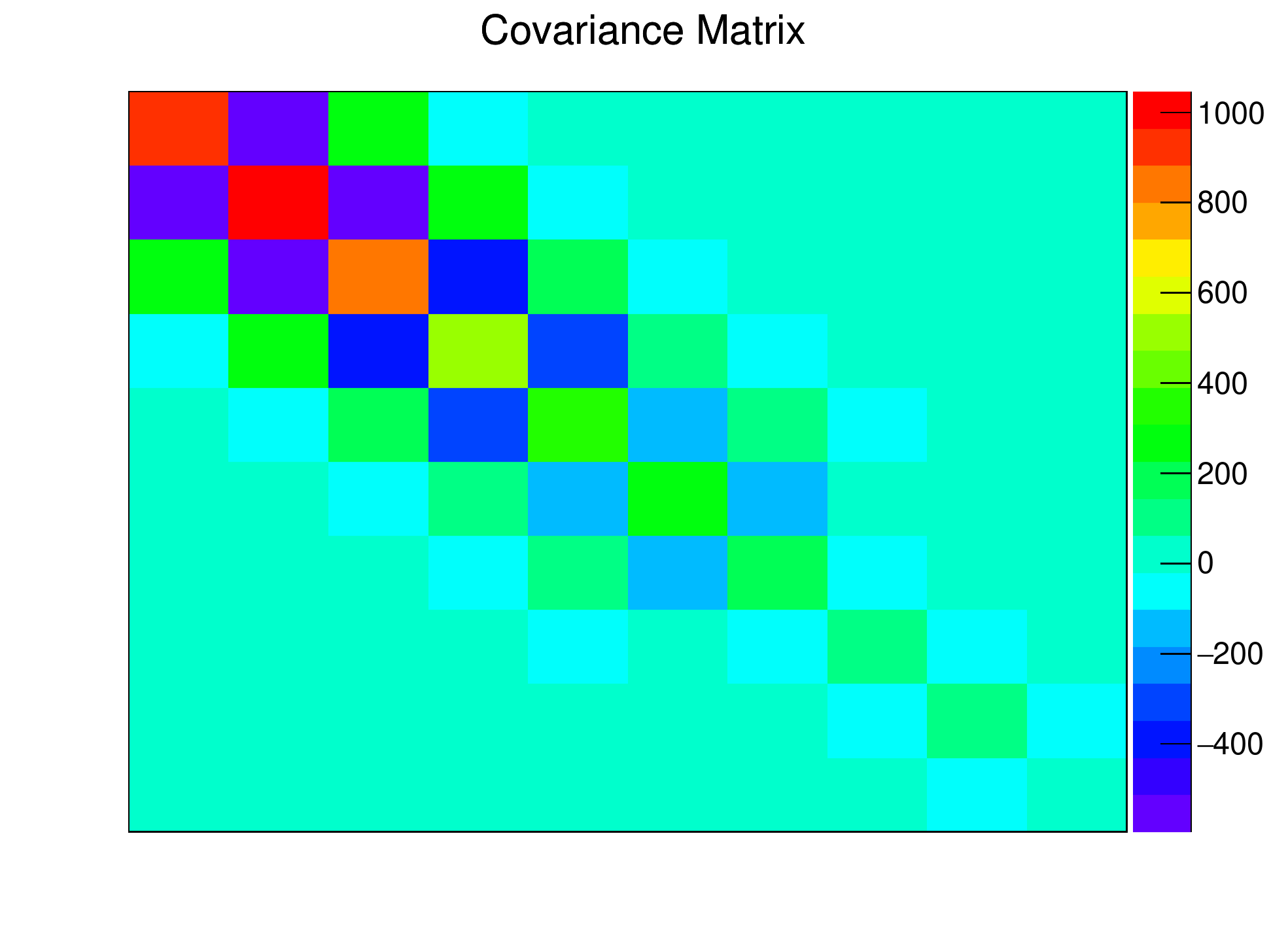}
\includegraphics[width=0.49\textwidth]{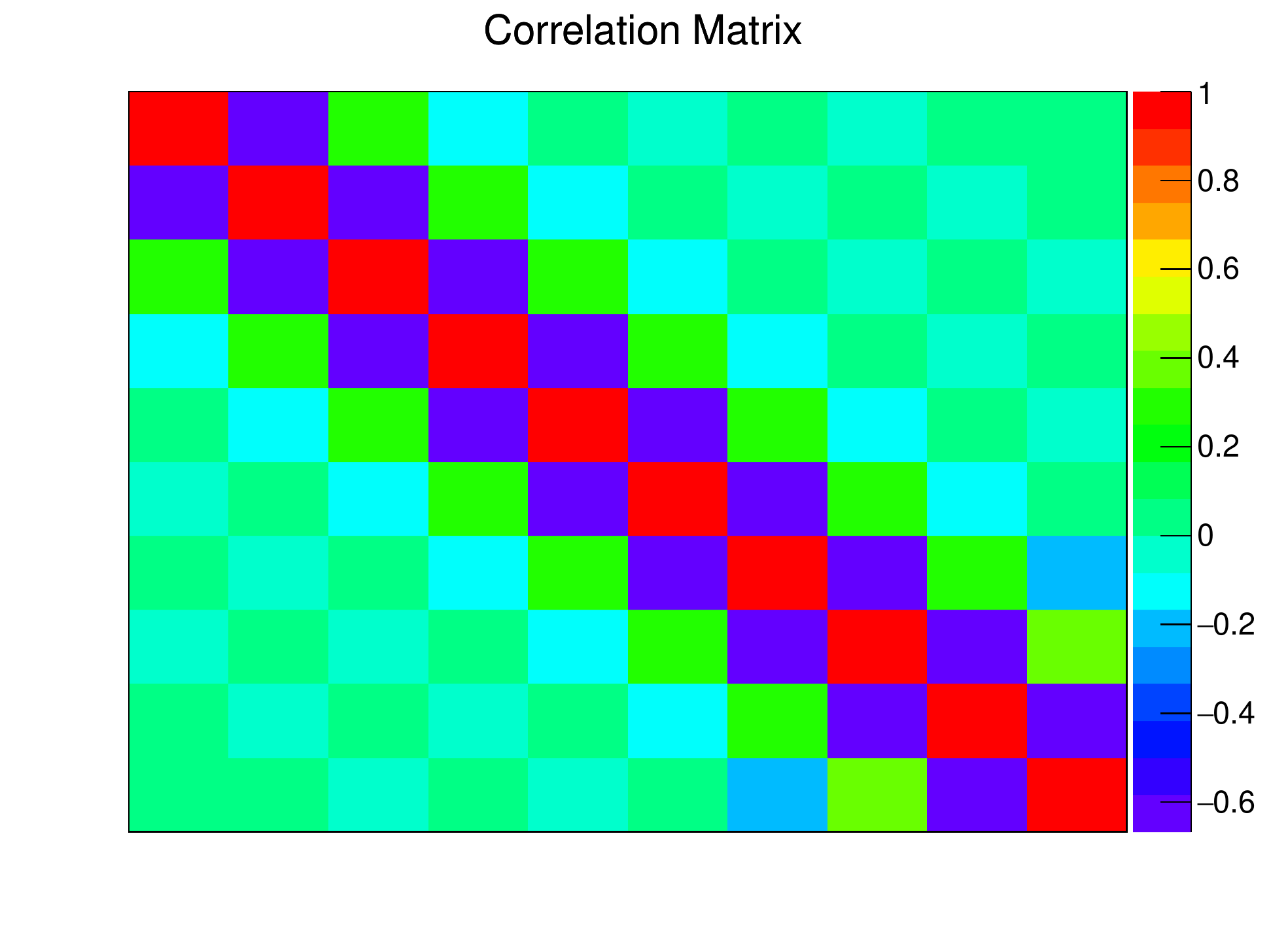}
\caption{(left) covariance matrix $\hatcovmu$ for unfolded estimates,
  as provided by the ML estimates (matrix inversion).  (right) The
  correlation matrix corresponding to $\hatcovmu$, with elements
  $\hatcovmu_{ij}/\sqrt{\hatcovmu_{ii}\hatcovmu_{jj}}$.
%----------- Matrices\_BinPurity\_Invert 1 2 
}
\label{matricesinvert}
\end{center}
\end{figure}

\begin{figure}
\begin{center}
\includegraphics[width=0.49\textwidth]{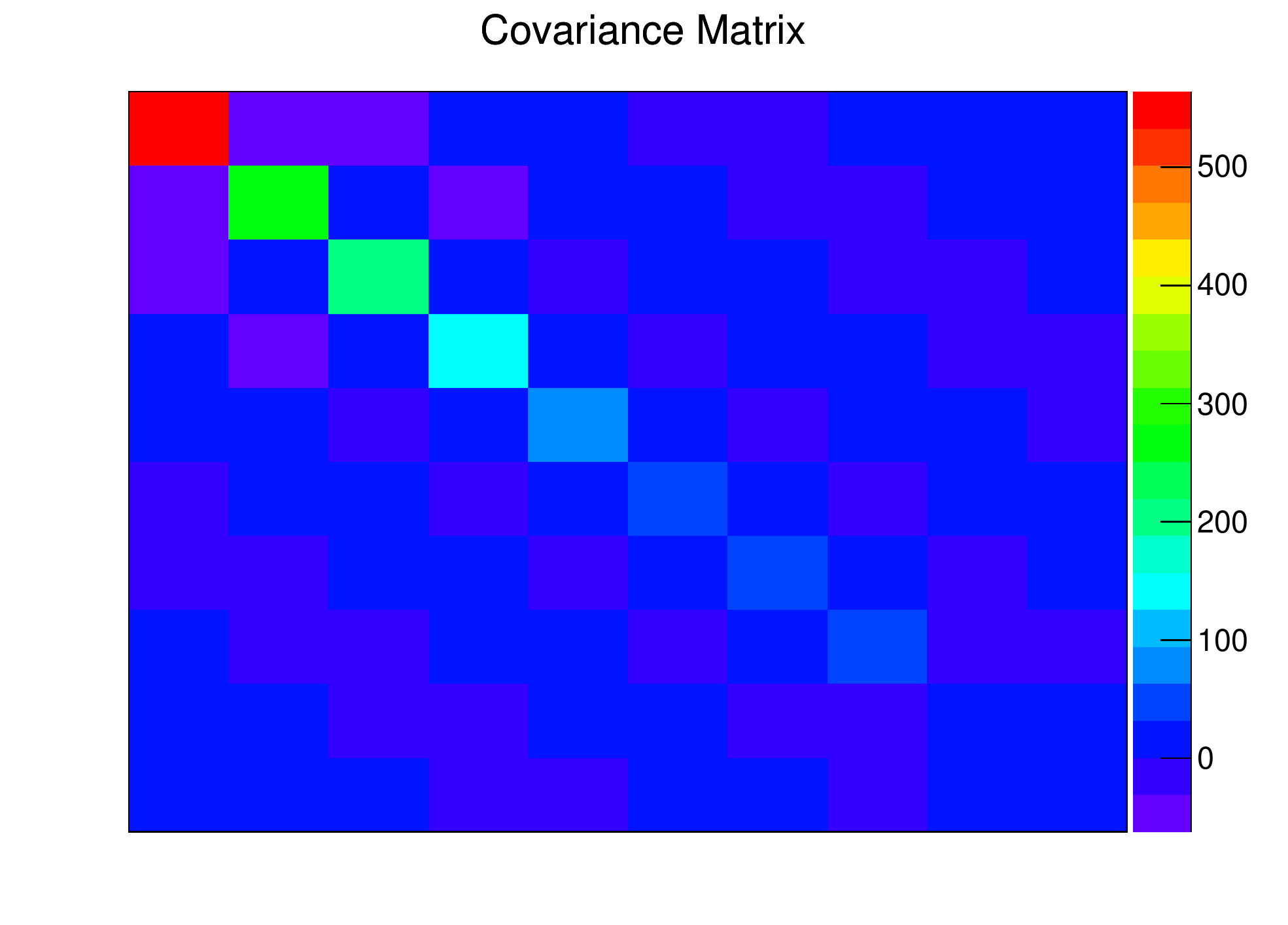}
\includegraphics[width=0.49\textwidth]{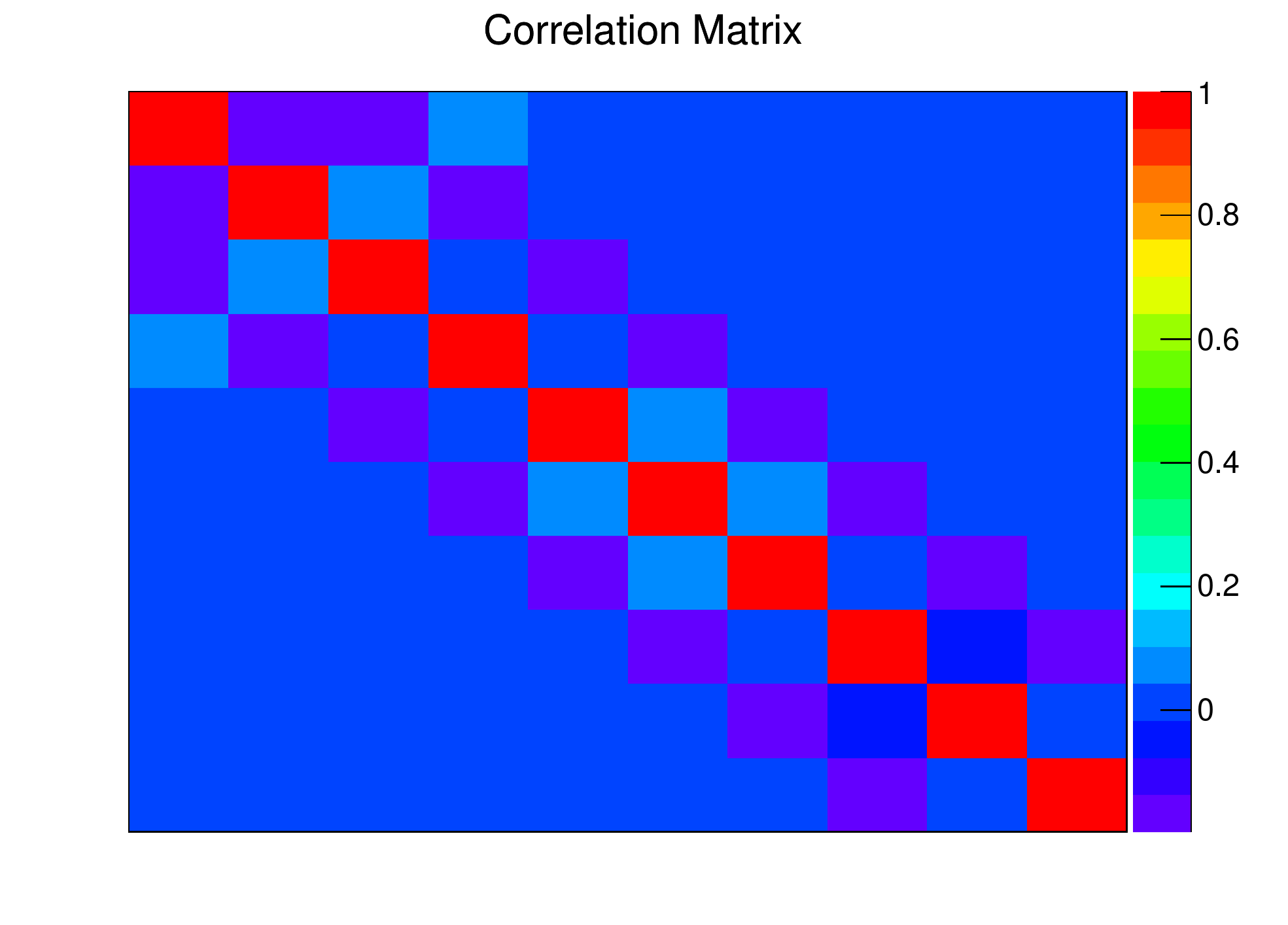}
\caption{(left) covariance matrix $\hatcovmu$ for unfolded estimates,
  as provided by the default iterative EM method.  (right) The
  correlation matrix corresponding to $\hatcovmu$, with elements
  $\hatcovmu_{ij}/\sqrt{\hatcovmu_{ii}\hatcovmu_{jj}}$.
%----------- Matrices\_BinPurity\_EM 1 2
}
\label{matricesiterative}
\end{center}
\end{figure}

Figure~\ref{histos} shows (in addition to the solid histograms
mentioned above) three points with error bars plotted in each bin,
calculated from a particular set of simulated data corresponding to
one experiment.  The three points are the bin contents when the
sampled values of $y$ and $x$ are binned, followed by that bin's
components of the set of unfolded estimates $\hatvecmu$.
Figure~\ref{matricesinvert}(left) shows the covariance matrix
$\hatcovmu$ for the estimates $\hatvecmu$ obtained for the same
particular simulated data set, unfolded by matrix inversion
(Eqn.~\ref{nuRmuinv}) to obtain the ML estimates.
Figure~\ref{matricesinvert} (right) shows the corresponding
correlation matrix with elements
$\hatcovmu_{ij}/\sqrt{\hatcovmu_{ii}\hatcovmu_{jj}}$.
Figure~\ref{matricesiterative} shows the corresponding matrices
obtained when unfolding by the iterative EM method with default number
of iterations.  For the ML solution, adjacent bins are negatively
correlated, while for the EM solution with default (4) iterations,
adjacent bins are positively correlated due to the implicit
regularization.

\begin{figure}
\begin{center}
\includegraphics[width=0.49\textwidth]{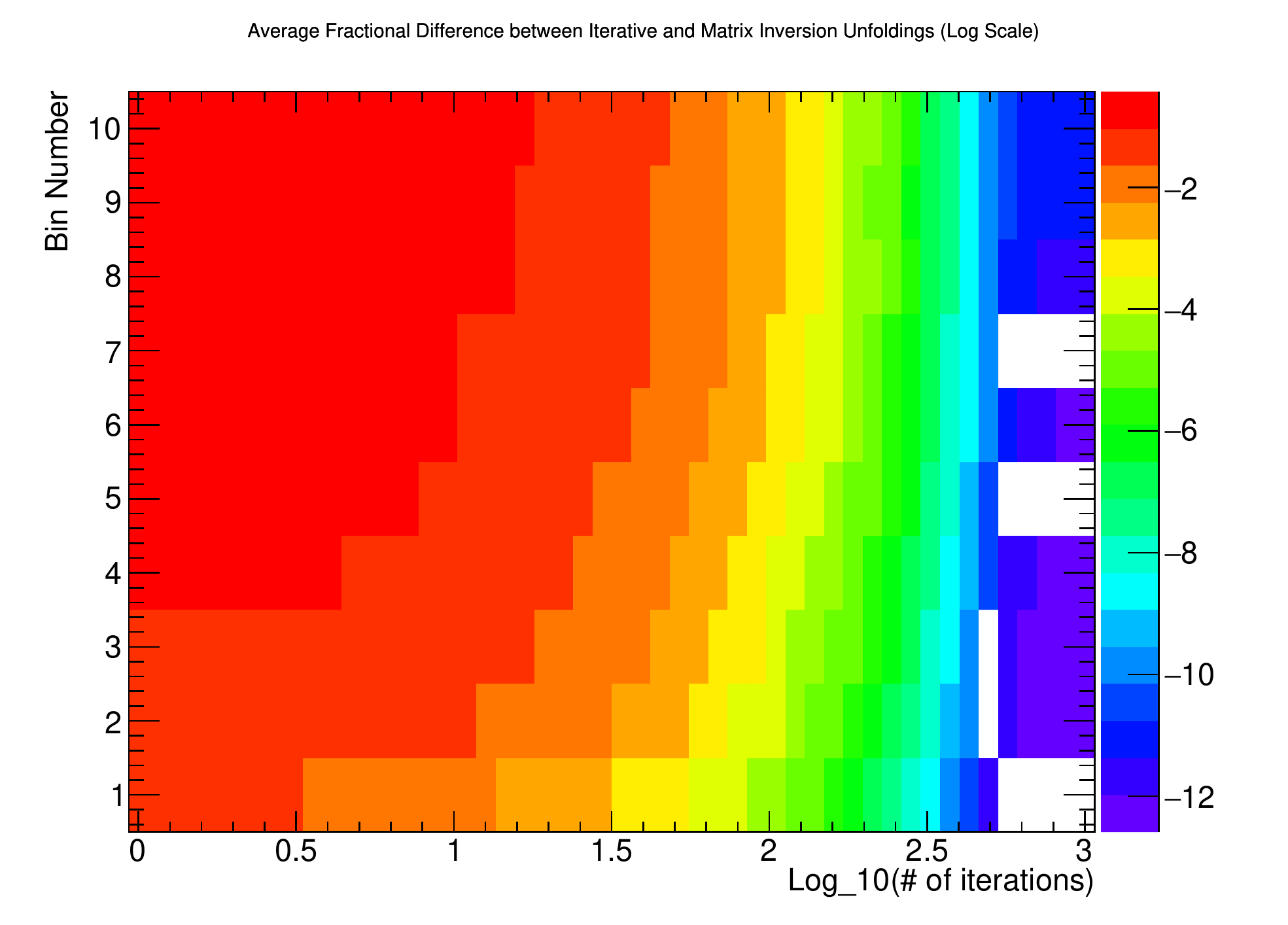}
\includegraphics[width=0.49\textwidth]{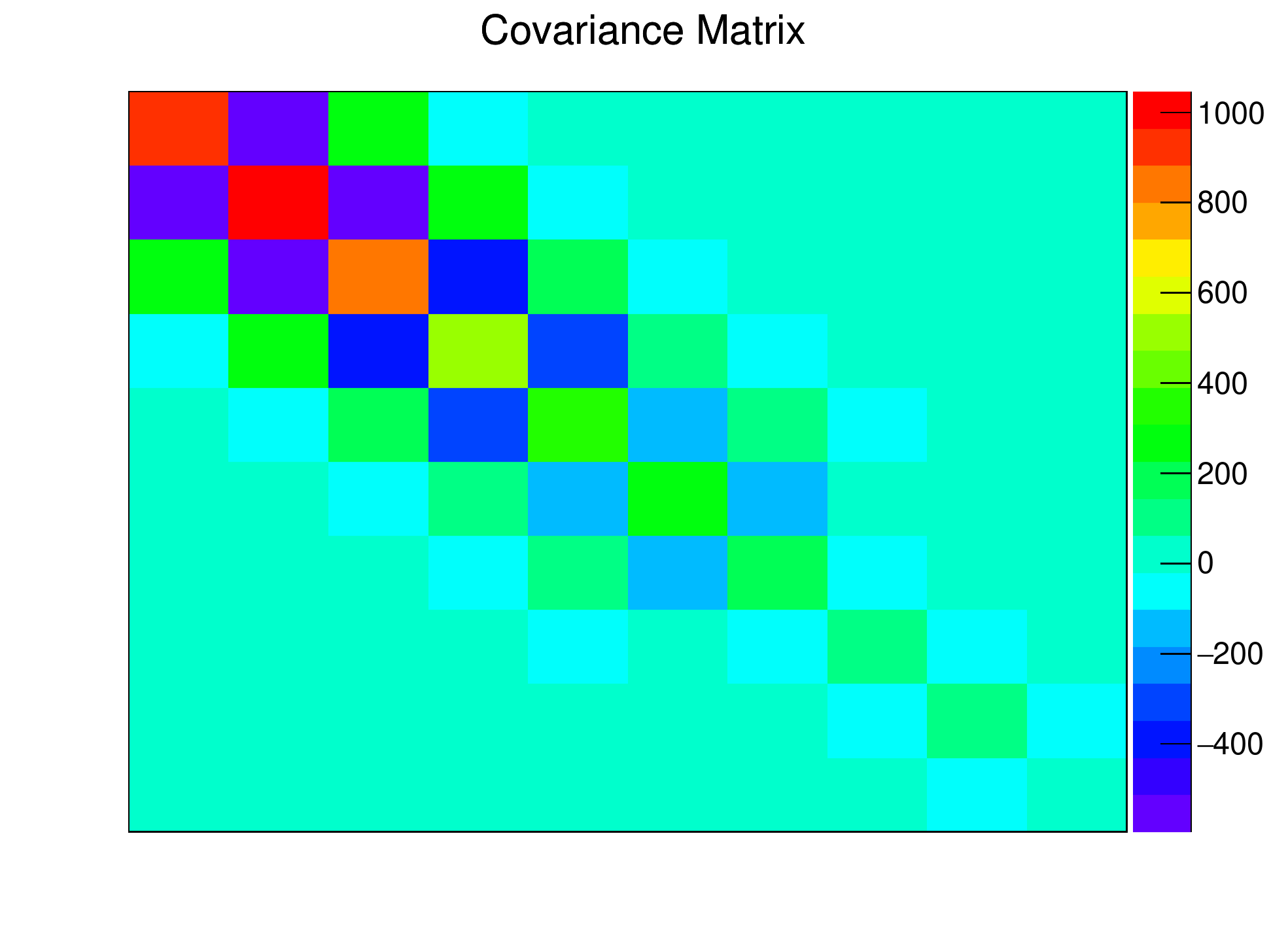}
\caption{(left) Fractional difference between the EM and ML solutions,
  for each of the ten histogram bins, as a function of the number of
  iterations. (right) The covariance matrix of the EM method after a
  large number of iterations, to be compared to the ML solution in
  Fig.~\ref{matricesinvert}(left).
%-----------IterativeConvergence.pdf
}
\label{converge}
\end{center}
\end{figure}

Figure~\ref{converge} shows an example of the convergence of iterative
EM unfolding to the ML solution for one simulated data set.  On the
left is the fractional difference between the EM and ML solutions, for
each of the ten histogram bins, as a function of the number of
iterations, reaching the numerical precision of the calculation. On
the right is the covariance matrix $\hatcovmu$ after a large number of
iterations, showing convergence to that obtained by matrix inversion
in Fig.~\ref{matricesinvert}(left).

\section{Hypothesis tests in the unfolded space}

Although the ML solution for $\hatvecmu$ may be difficult for a human
to examine visually, if the covariance matrix $\covmu$ is well enough
behaved, then a computer can readily calculate a chisquare GOF test
statistic in the unfolded space by using the generalization of
Eqn.~\ref{chisq}, namely the usual formula for GOF of Gaussian
measurements with correlations~\cite{PDG},
\begin{equation}
\label{chicorr}
\chicorr = (\hatvecmu - \vec\mu)^T\,\covmu^{-1}\,(\hatvecmu - \vec\mu).
\end{equation}

If unfolding is performed by matrix inversion (when equal to the ML
solution), then substituting $\hatvecmu = R^{-1}\,\vec n$ from
Eqn.~\ref{nuRmuinv}, $\vec\mu = R^{-1}\,\vec\nu$ from
Eqn.~\ref{nuRmu}, and $\covmu^{-1} = R^T\,V^{-1}\,R$ from
Eqn.~\ref{covmu}, yields
\begin{equation}
\label{chicorrinv}
\chicorr = (\vec n - \vec\nu)^T\,V^{-1}\,(\vec n - \vec\nu).
\end{equation}
So for $V_{ij} = \delta_{ij}\nu_i$ as assumed by Cowan, this
$\chicorr$ calculated in the unfolded space is equal to Pearson's
chisquare (Eqn.~\ref{pearson}) in the smeared space.

If however one substitutes $\hatvecnu = \vec n$ for $\vec\nu$ as in
Eqn.~\ref{nun}, then $\chicorr$ in the unfolded space is equal to
Neyman's chisquare in the smeared space!  This is the case in the
implementation of RooUnfold that we are using, as noted below in the
figures.

\begin{figure}
\begin{center}
\includegraphics[width=0.49\textwidth]{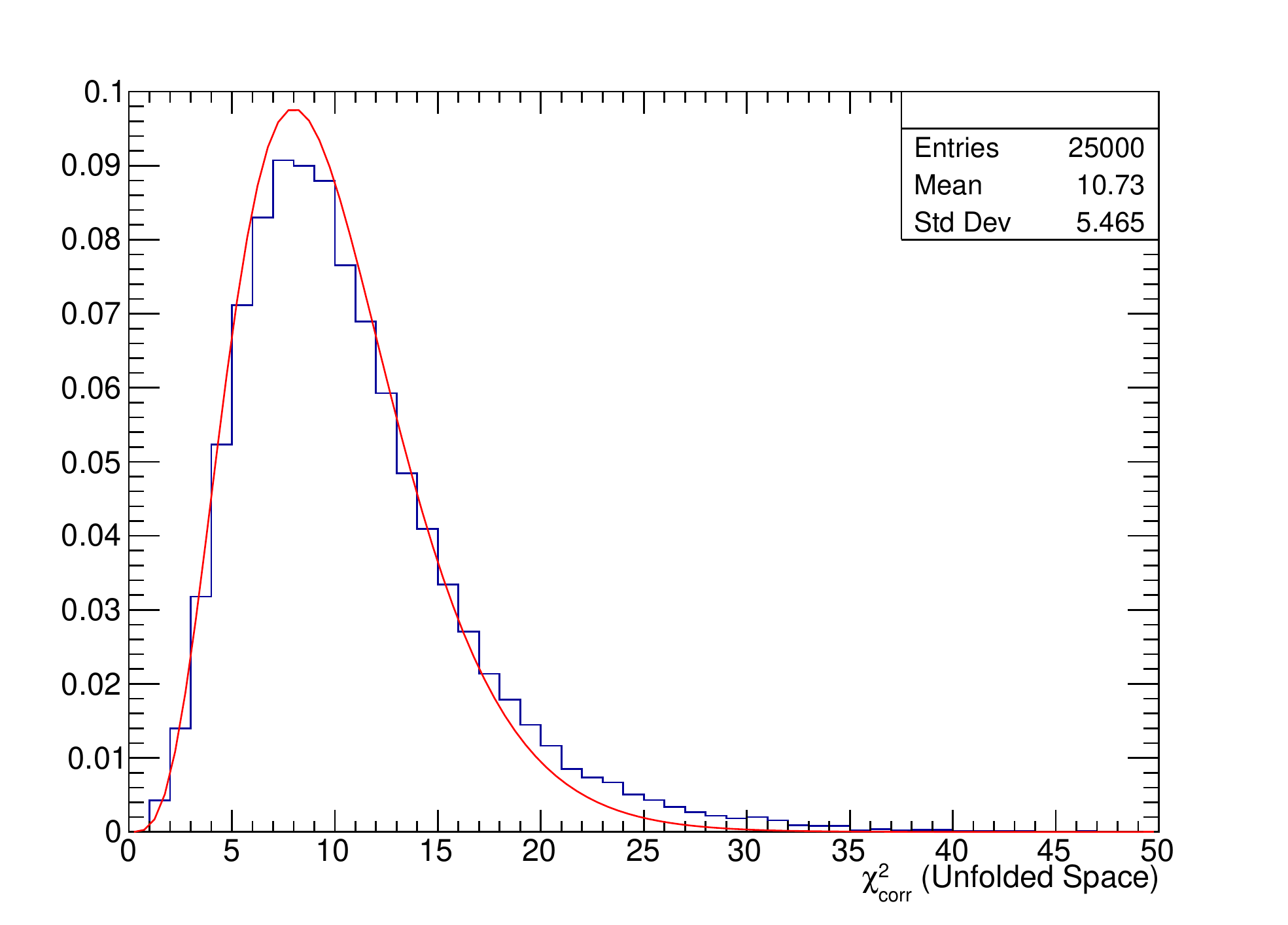}
\includegraphics[width=0.49\textwidth]{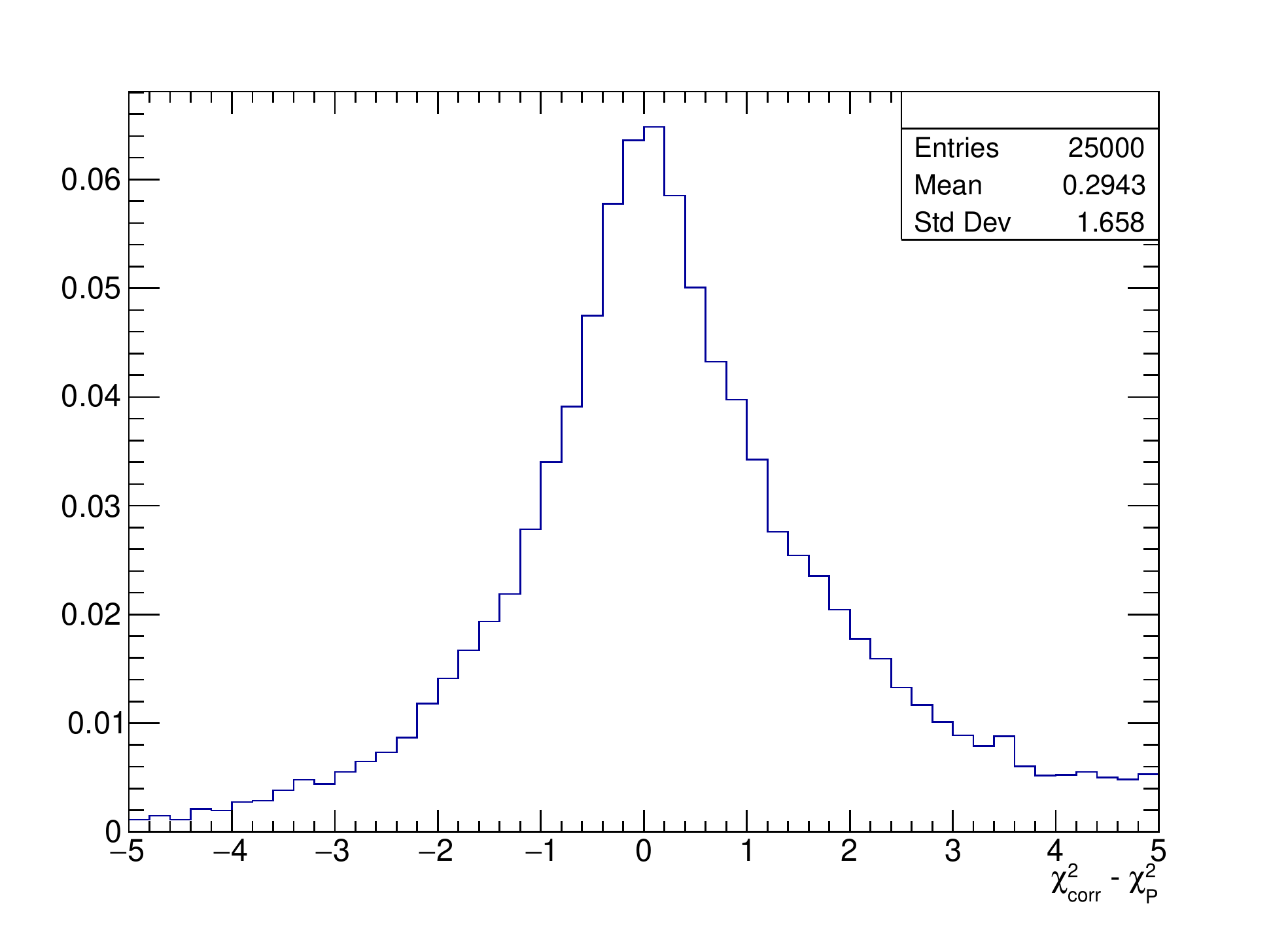}
\includegraphics[width=0.49\textwidth]{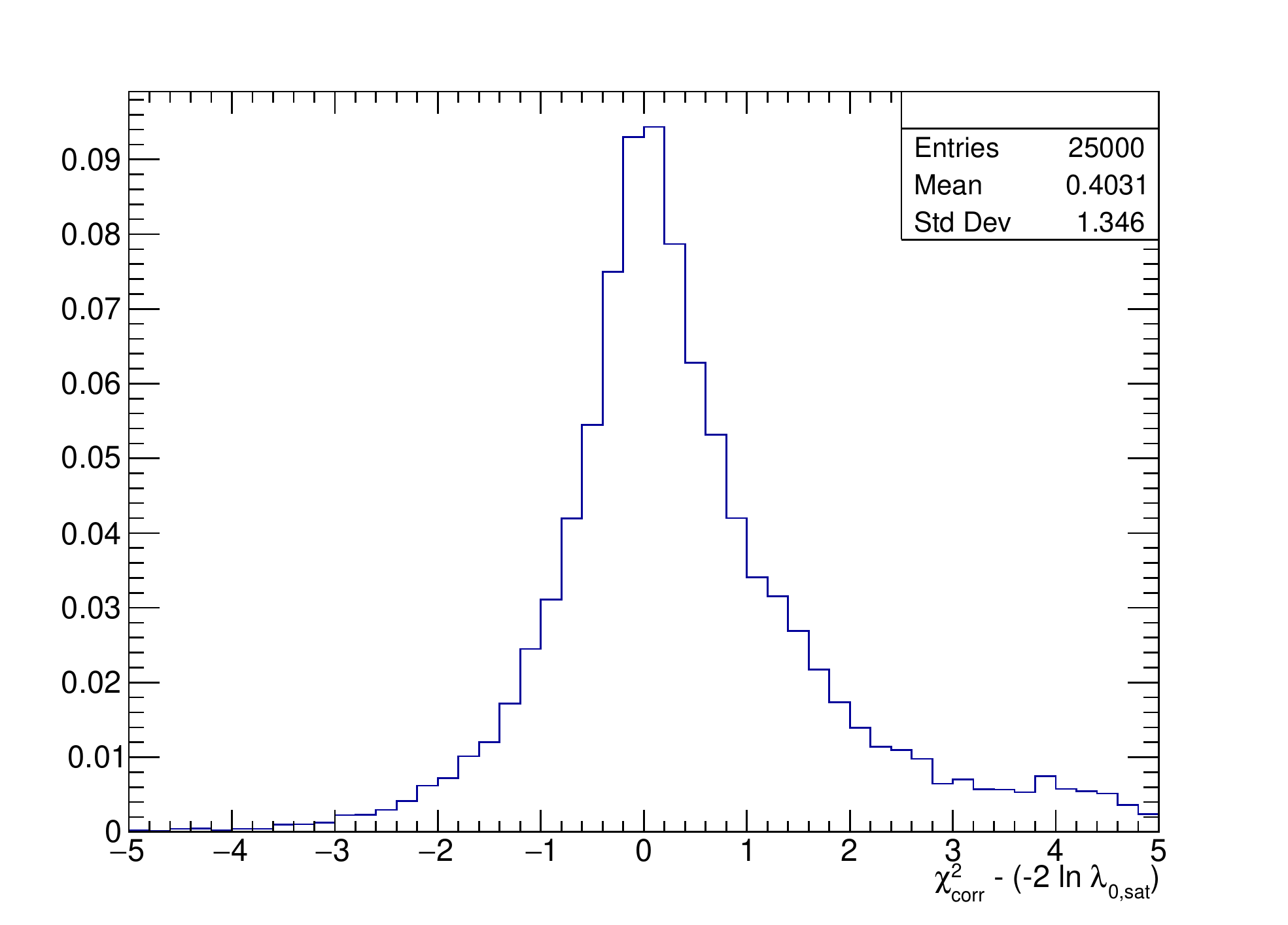}
\caption{Quantities calculated after unfolding using ML
  estimates. (top left) Histograms of generalized GOF test statistic
  $\chicorr$ that tests for compatibility with $H_0$ in the unfolded
  space, for the same events generated under $H_0$ as those used in
  the smeared-space test of Fig.~\ref{nullgofsmeared}.  (top right)
  For these events, histogram of the difference between $\chicorr$ in
  the unfolded space and $\chip$ in the smeared space.  (bottom) For
  these events, histogram of the difference between $\chicorr$ in the
  unfolded space and the GOF test statistic $-2\ln\lambda_{0,sat}$ in
  the smeared space.
%---------- OneHypothesis\_1DSlice\_Invert 3 5 6
}
\label{nulgofunfoldedinvert}
\end{center}
\end{figure}

\begin{figure}
\begin{center}
\includegraphics[width=0.49\textwidth]{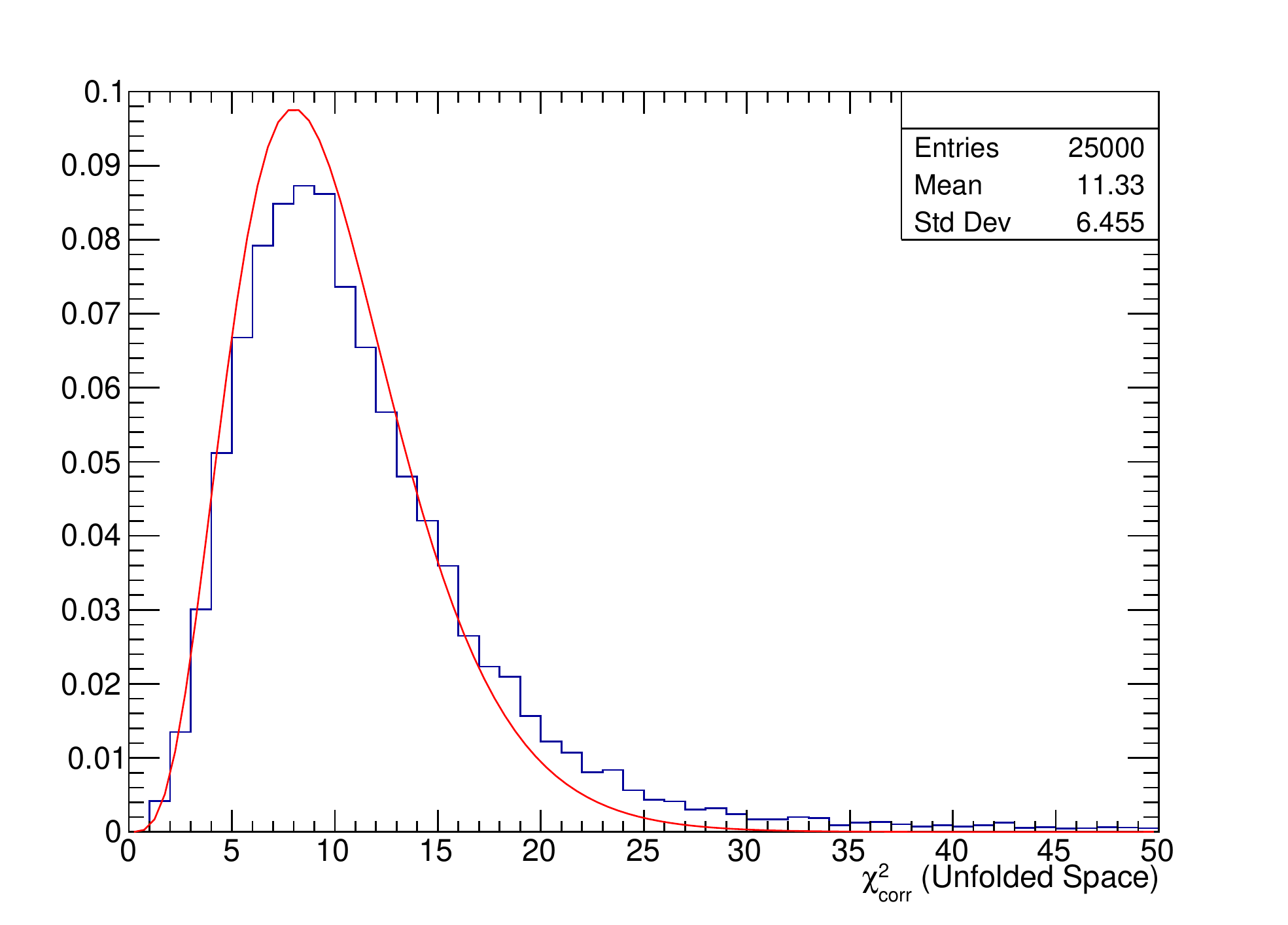}
\includegraphics[width=0.49\textwidth]{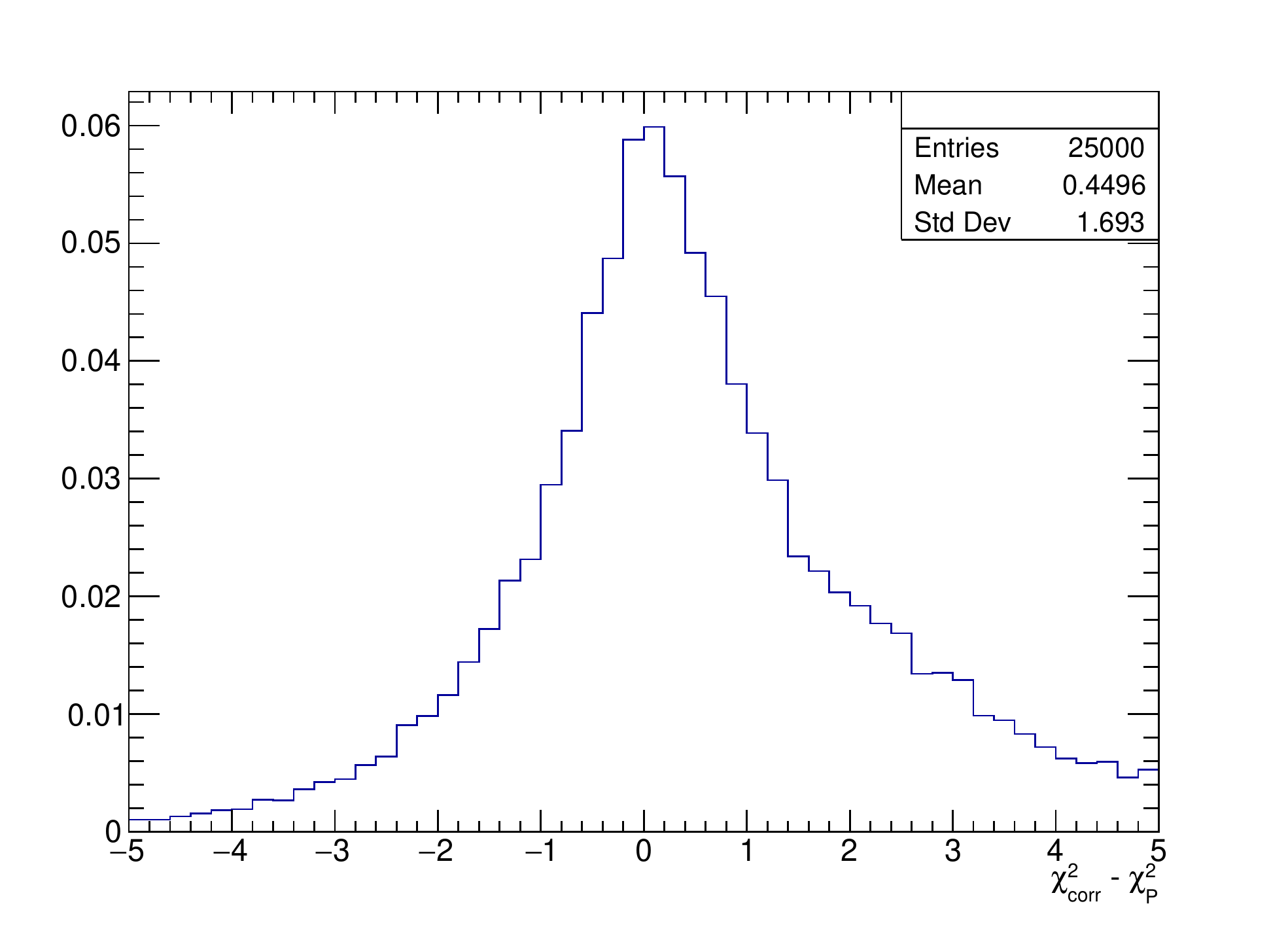}
\includegraphics[width=0.49\textwidth]{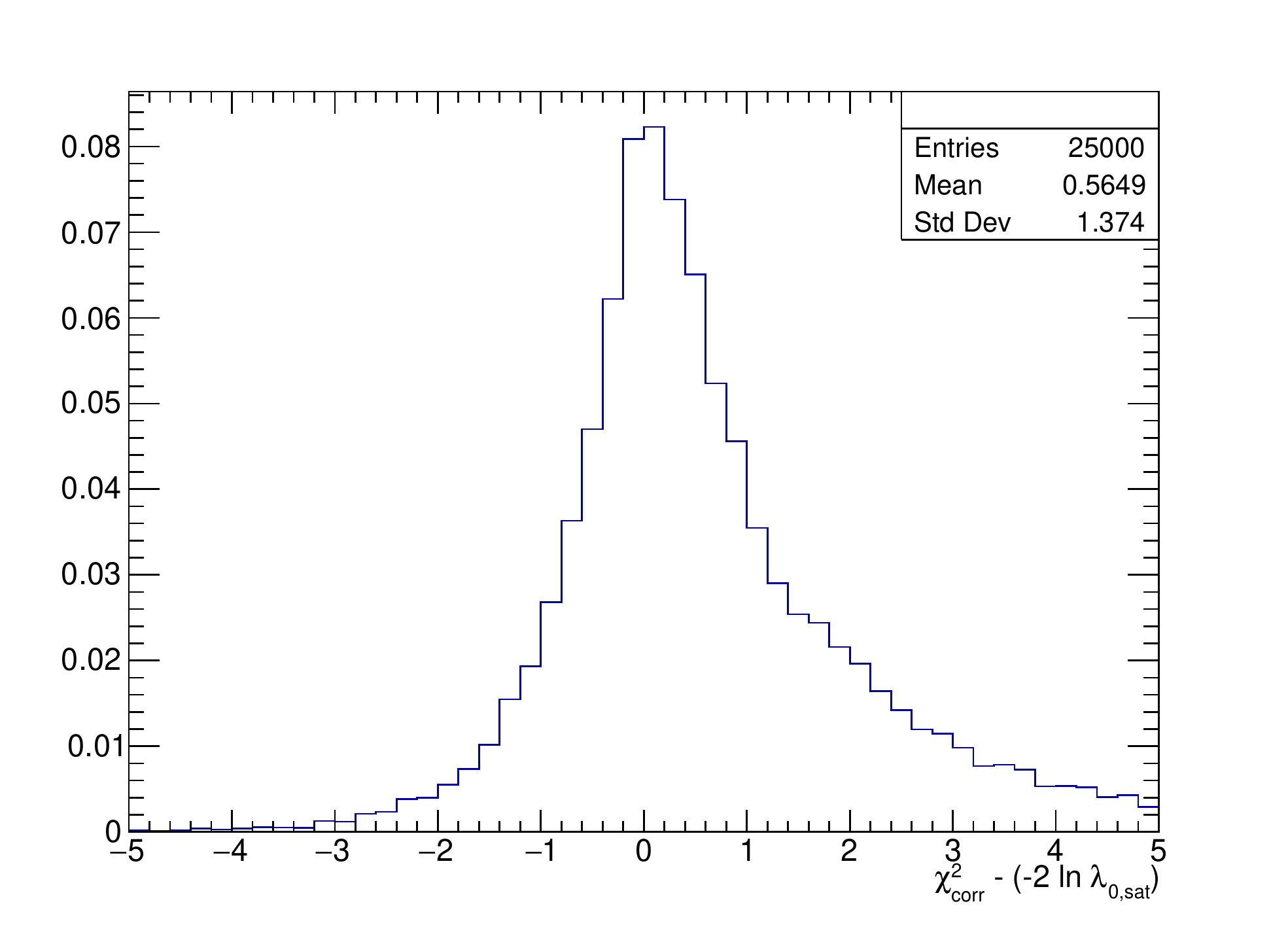}
\caption{The same quantities as in Fig.~\ref{nulgofunfoldedinvert},
  here calculated after unfolding using the iterative EM method with
  default (four) iterations.
%---------- OneHypothesis\_1DSlice 3 5 6  
}
\label{nulgofunfolded}
\end{center}
\end{figure}

For events unfolded with the ML estimates,
Figure~\ref{nulgofunfoldedinvert} (top left) shows the results of such
a $\chicorr$ GOF test with respect to the null hypothesis using same
events used in Fig.~\ref{nullgofsmeared}. As foreseen, the histogram
is identical (apart from numerical artifacts) with the histogram of
$\chin$ in Fig.~\ref{nullgofsmeared} (bottom left).
Figure~\ref{nulgofunfoldedinvert} (top right) show the event-by-event
difference of $\chicorr$ and Pearson's $\chi^2$ in the smeared space,
and Figure~\ref{nulgofunfoldedinvert} (bottom) is the difference with
respect to $-2\ln\lambda_{0,{\rm sat}}$ in the smeared space.
Figure~\ref{nulgofunfolded} shows the same quantities calculated after
unfolding using the iterative EM method with default iterations.

For these tests using ML unfolding, the noticeable difference between
the GOF test in the smeared space with that in the unfolded space is
directly traced to the fact that the test in the unfolded space is
equivalent to $\chin$ in the smeared space, which is an inferior GOF
test compared to the likelihood ratio test statistic
$-2\ln\lambda_{0,sat}$.  It seems remarkable that, even though
unfolding by matrix inversion would appear not to lose information, in
practice the way the information is used (linearizing the problem via
expressing the result via a covariance matrix) already results in some
failures of the bottom-line test of GOF.  This is without any
regularization or approximate EM inversion.

\begin{figure}
\begin{center}
\includegraphics[width=0.49\textwidth]{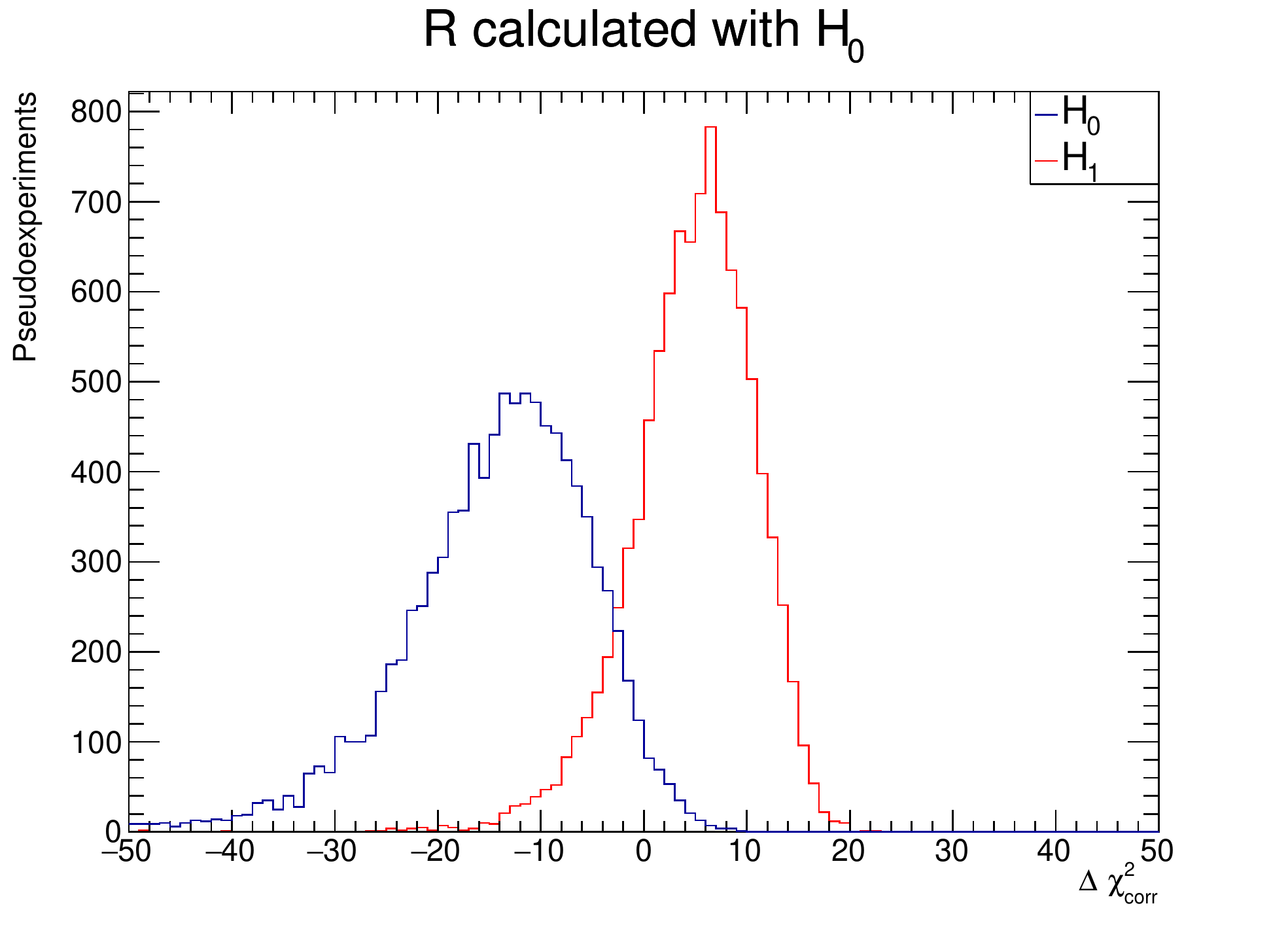}
\includegraphics[width=0.49\textwidth]{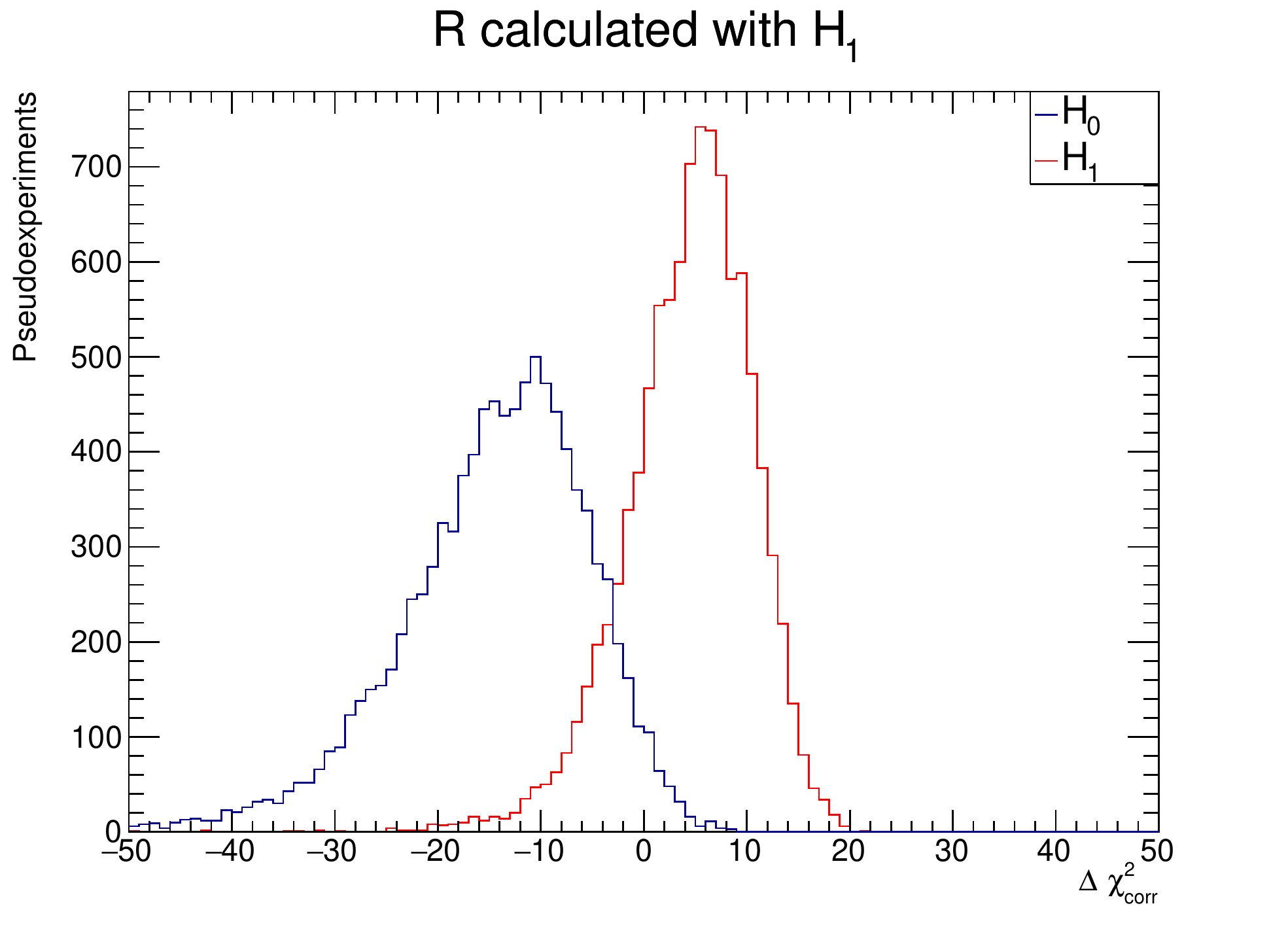}
\caption{ (left) For the same events as those used in
  Fig.~\ref{lambdah0h1}, histogram of the test statistic
  $\Delta\chicorr$ in the unfolded space, for events generated under
  $H_0$ (in blue) and $H_1$ (in red), with $R$ calculated using
  $H_0$. (right) For the same events, histograms of the test statistic
  $\Delta\chicorr$ in the unfolded space, with $R$ calculated using
  $H_1$.
%--------- HypothesisComparison 2 3
}
\label{delchi}
\end{center}
\end{figure}

\clearpage
For the histogram of each simulated experiment, the GOF statistic
$\chicorr$ is calculated with respect to the prediction of $H_0$ and
also with respect to the prediction of $H_1$.  The difference of these
two values, $\Delta\chicorr$, is then a test statistic for testing
$H_0$ vs.\ $H_1$, analogous to the test statistic
$-2\ln\lambda_{0,1}$.  Figure~\ref{delchi} shows, for the same events
as those used in Fig.~\ref{lambdah0h1}, histograms of the test
statistic $\Delta\chicorr$ in the unfolded space for events generated
under $H_0$ and under $H_1$, with $R$ calculated using $H_0$ and using
$H_1$. For the default problem studied here, the dependence on $R$ is
not large.  Thus unless otherwise specified, all other plots use $R$
calculated under $H_0$.

\begin{figure}
\begin{center}
\includegraphics[width=0.49\textwidth]{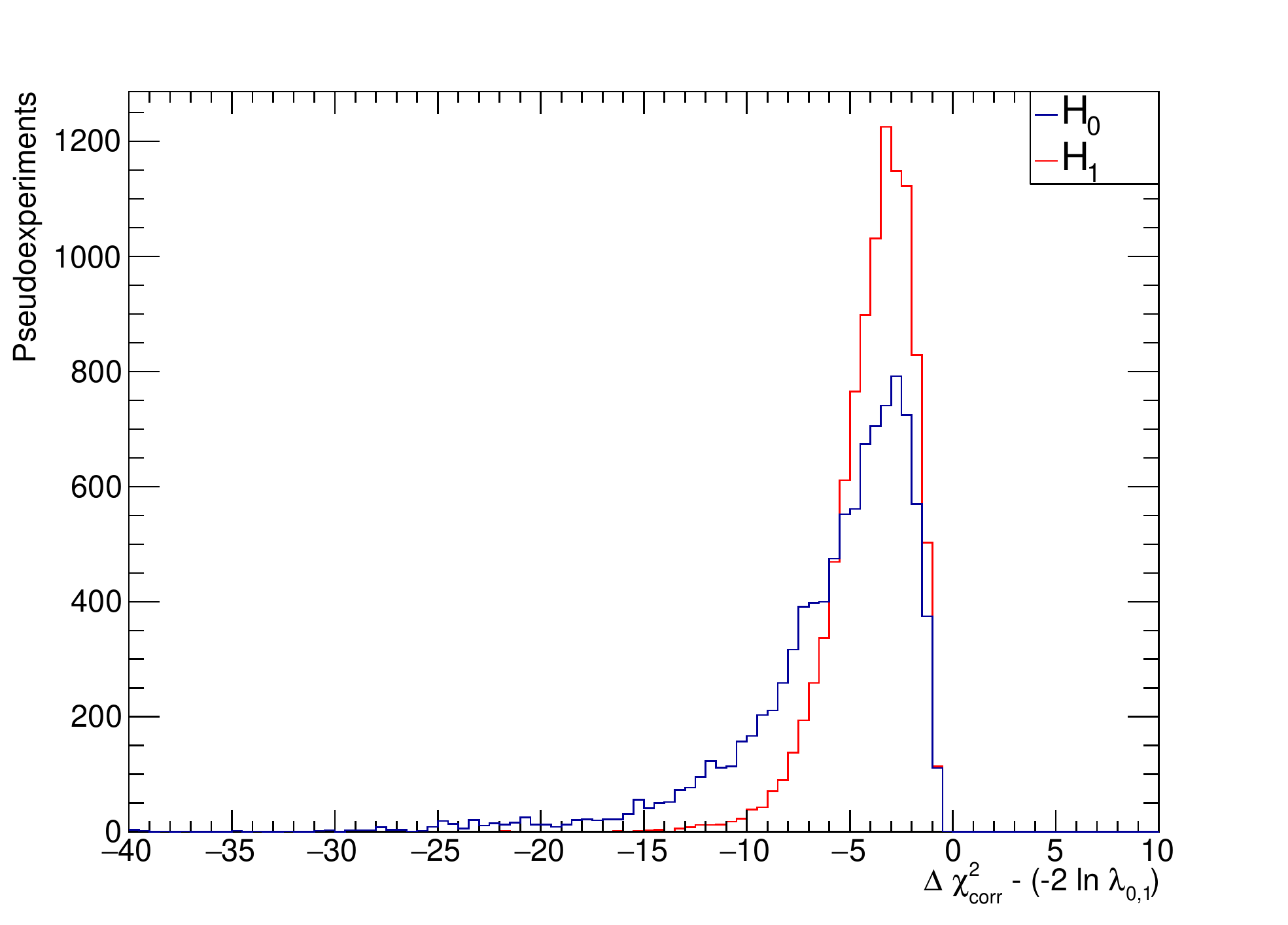}
\includegraphics[width=0.49\textwidth]{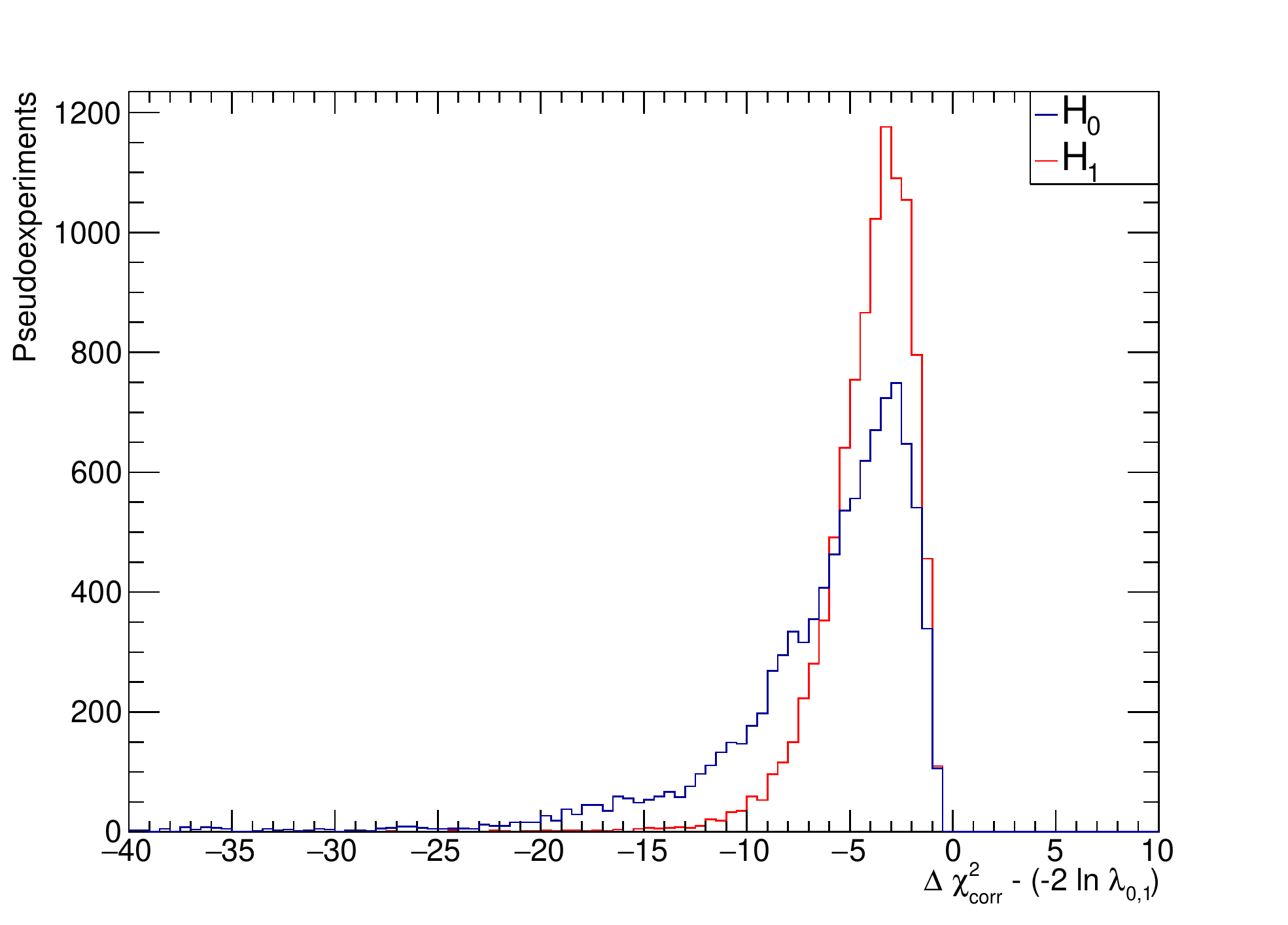}
\caption{For the events in Figs.~\ref{lambdah0h1} and in
  \ref{delchi}(left), histogram of the event-by-event difference of
  $-2\ln\lambda_{0,1}$ and $\Delta\chicorr$.  In the left histogram,
  ML unfolding is used, while in the right histogram, iterative EM
  unfolding is used.
%--------HypothesisComparison\_Invert pages 4
%--------HypothesisComparison pages 4
}
\label{deldel}
\end{center}
\end{figure}

\clearpage
Figure~\ref{deldel} shows, for the events in Figs.~\ref{lambdah0h1}
and in \ref{delchi}, histograms of the event-by-event difference of
$-2\ln\lambda_{0,1}$ and $\Delta\chicorr$. The red curves correspond
to events generated under $H_0$, while the blue curves are for events
generated under $H_1$.  The unfolding method is ML on the left and
iterative EM on the right.  This is an example of a {\em bottom-line
  test}: does one obtain the same answers in the smeared and unfolded
spaces?  There are differences apparent with both unfolding
techniques.  Since the events generated under both $H_0$ and $H_1$ are
shifted in the same direction, the full implications are not
immediately clear.  Thus we turn to ROC curves or equivalent curves
from Neyman-Pearson hypothesis testing.

\begin{figure}
\begin{center}
\includegraphics[width=0.49\textwidth]{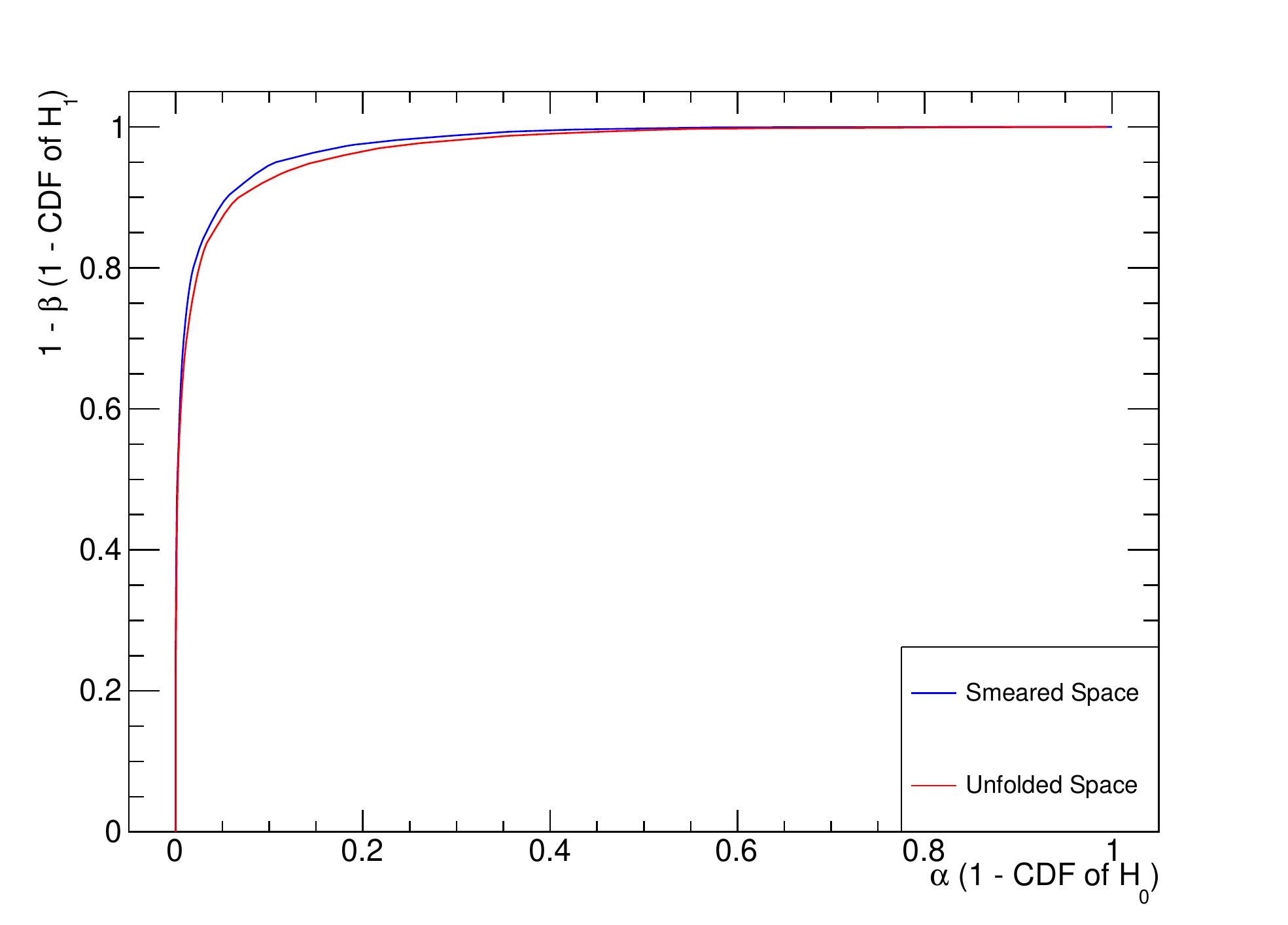}
\includegraphics[width=0.49\textwidth]{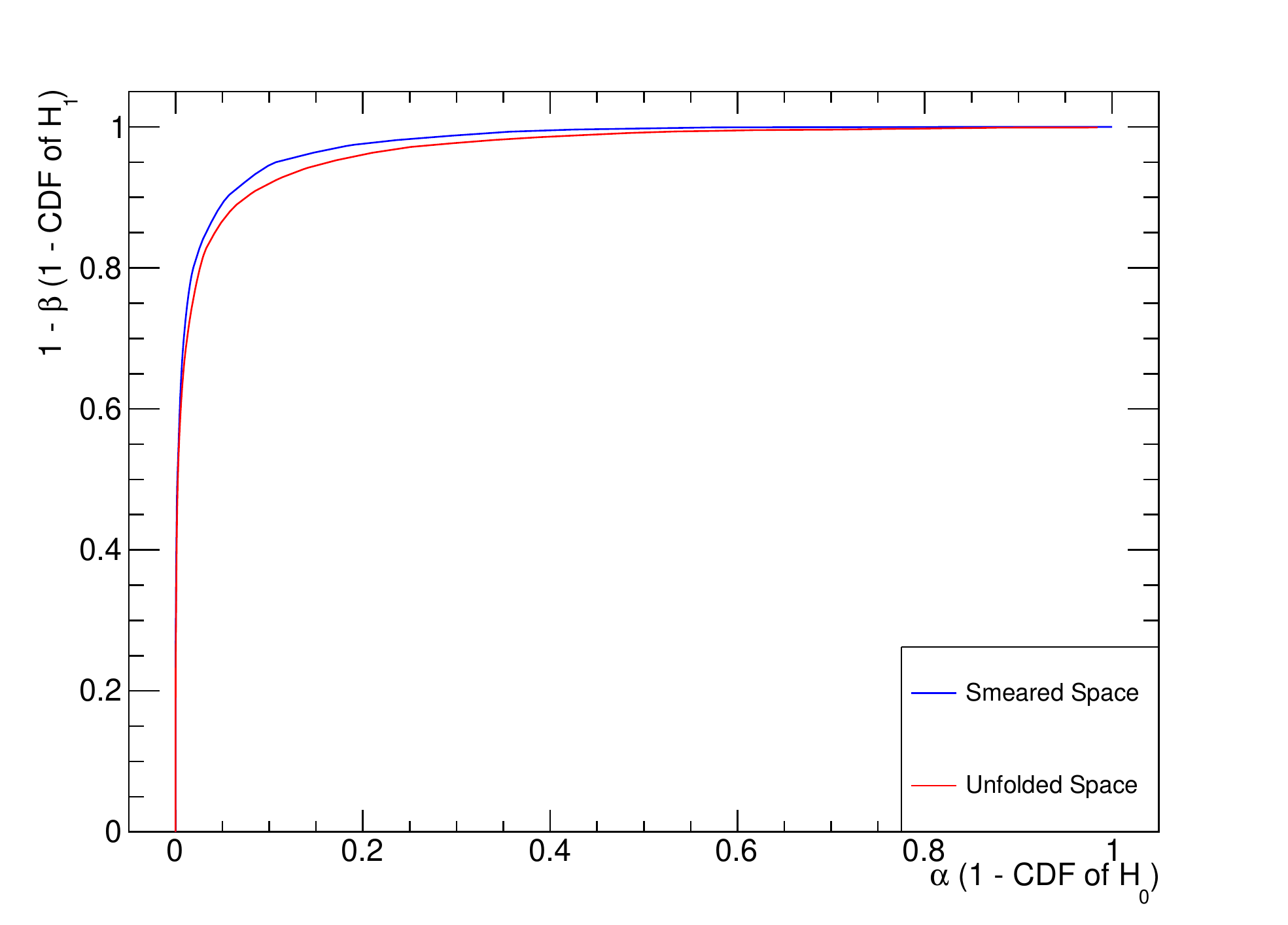}
\caption{For the events in Figs.~\ref{lambdah0h1} and
  \ref{delchi}(left), ROC curves for classification performed in the
  smeared space (blue curve) and in the unsmeared space (red
  curve). (left) unfolding by ML, and (right) unfolding by iterative
  EM.
%--------HypothesisComparison\_Invert page 5
%--------HypothesisComparison page 5
}
\label{ROC}
\end{center}
\end{figure}

\begin{figure}
\begin{center}
\includegraphics[width=0.49\textwidth]{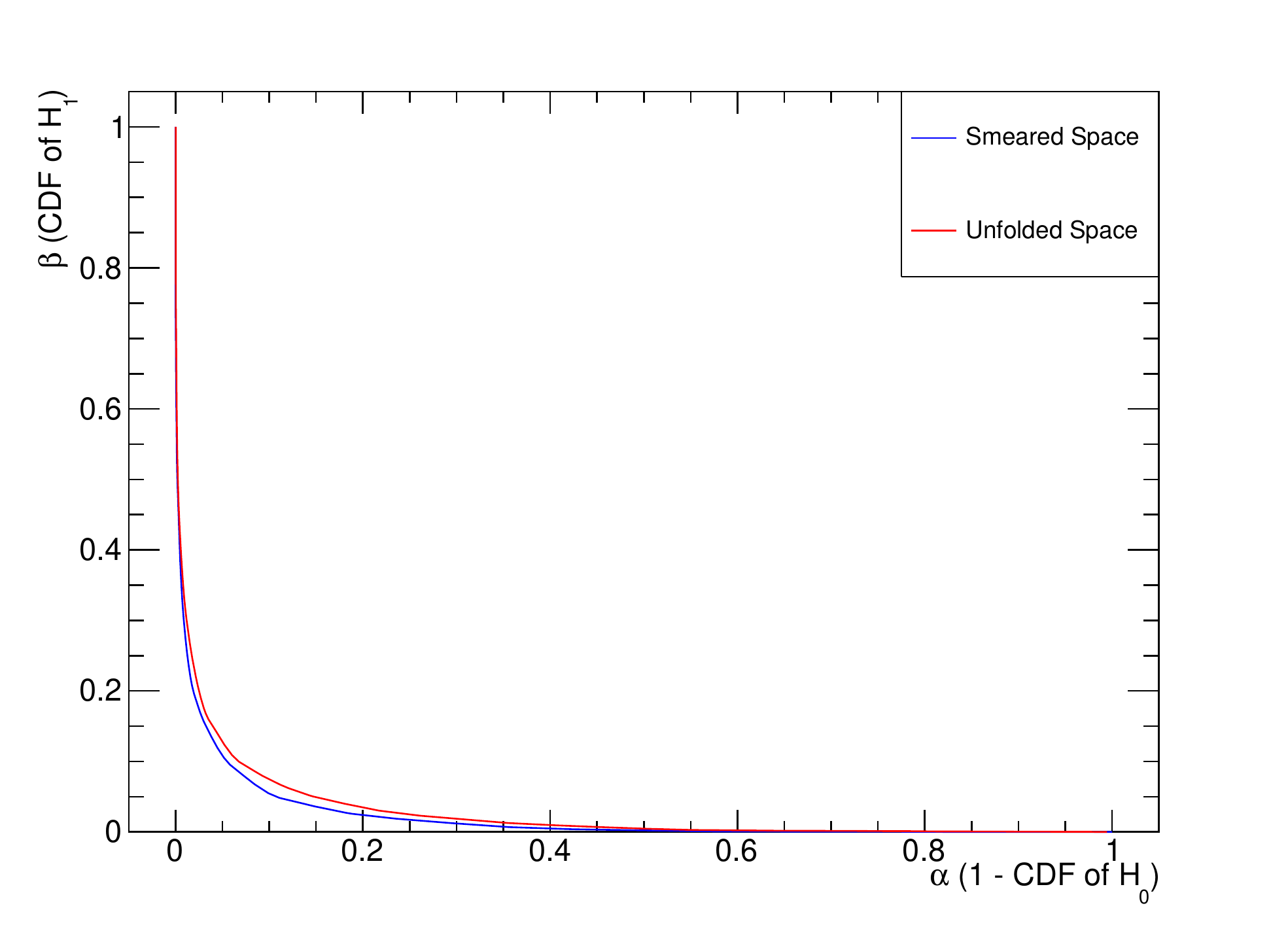}
\includegraphics[width=0.49\textwidth]{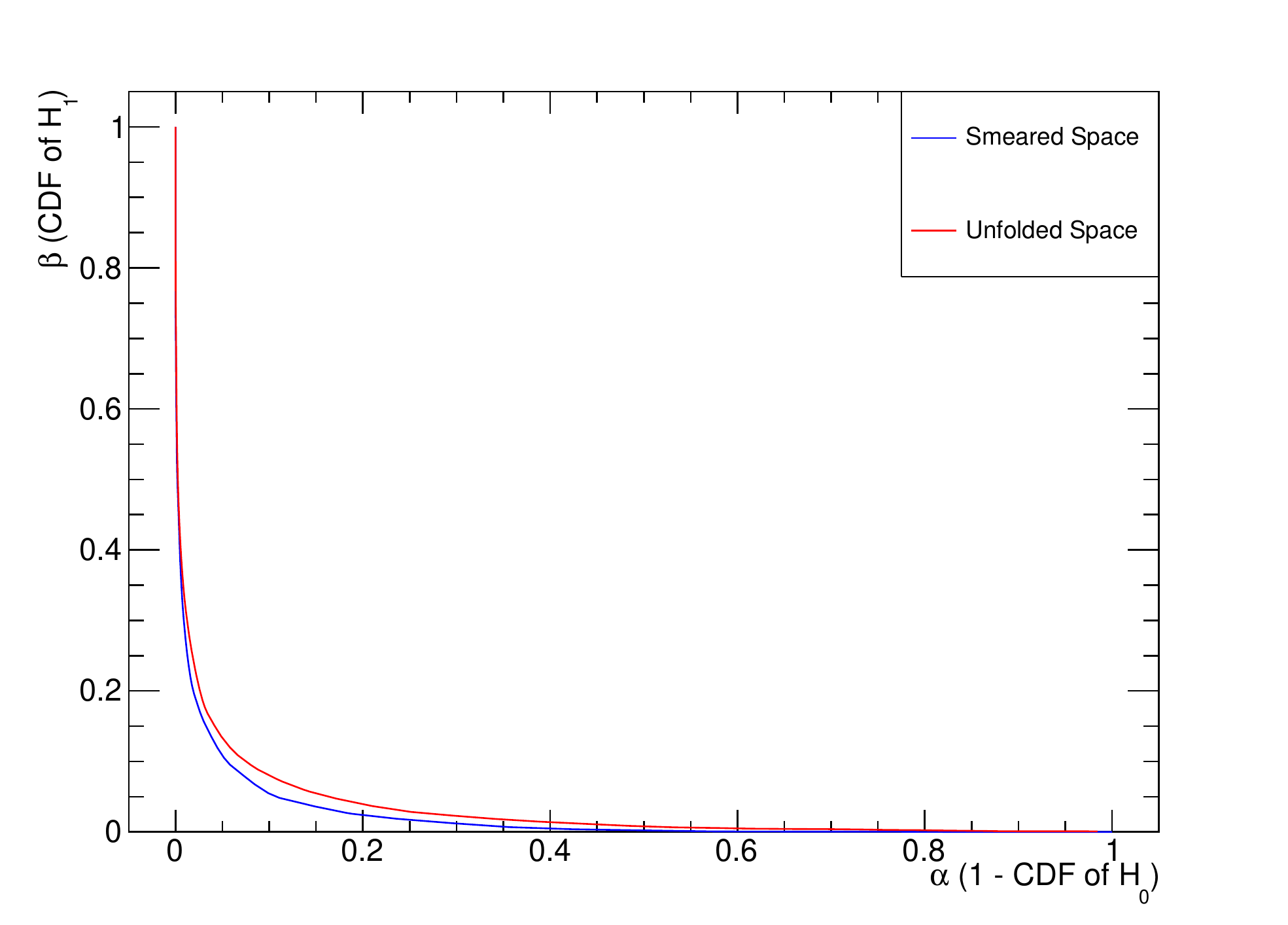}
\caption{For the events in Figs.~\ref{lambdah0h1} and
  \ref{delchi}(left), plots of $\beta$ vs. $\alpha$, for
  classification performed in the smeared space (blue curve) and in
  the unsmeared space (red curve). (left) unfolding by ML, and (right)
  unfolding by iterative EM.
%--------HypothesisComparison\_Invert page 6
%--------HypothesisComparison page 6
}
\label{alphabeta}
\end{center}
\end{figure}

\begin{figure}
\begin{center}
\includegraphics[width=0.49\textwidth]{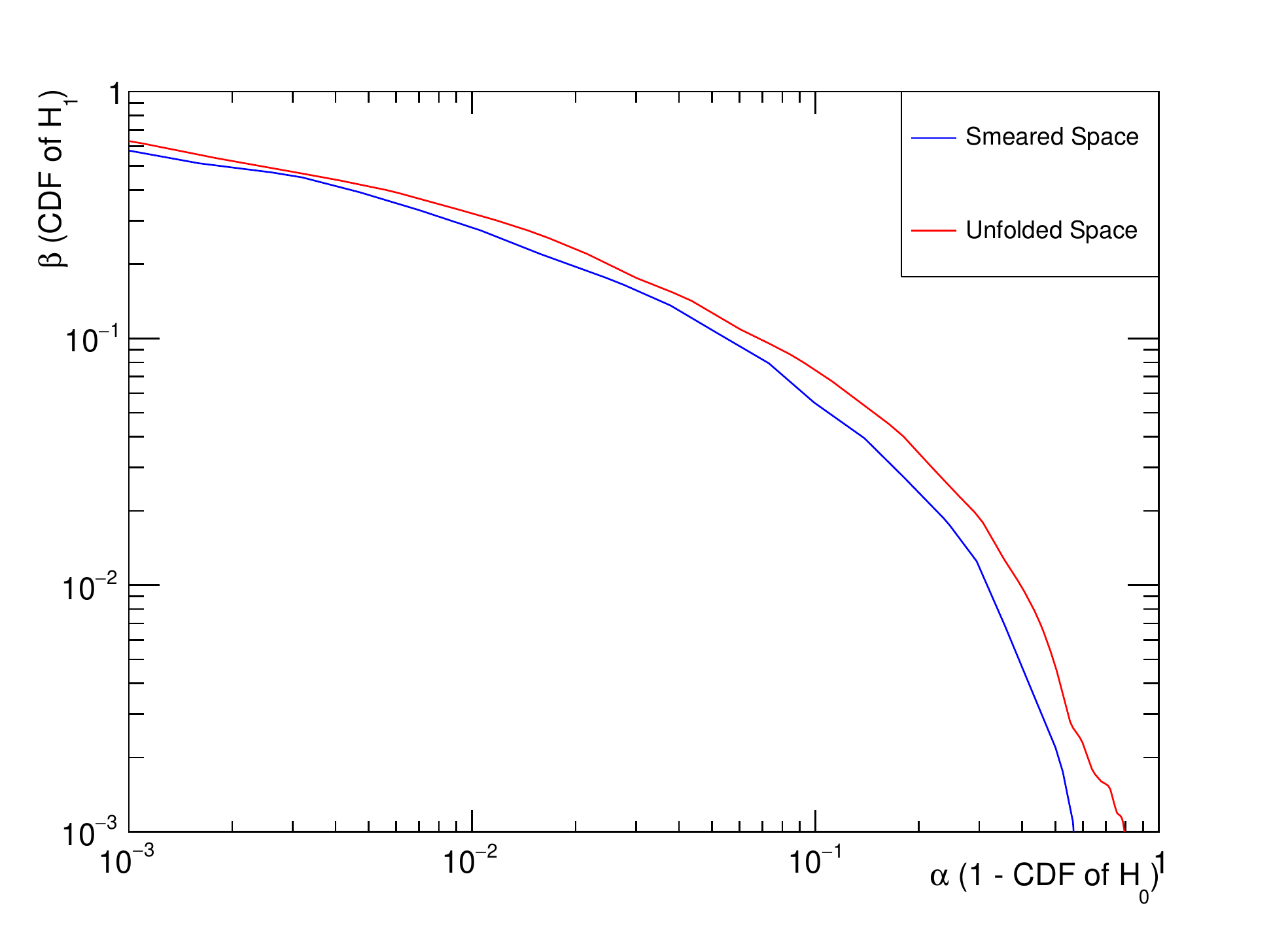}
\includegraphics[width=0.49\textwidth]{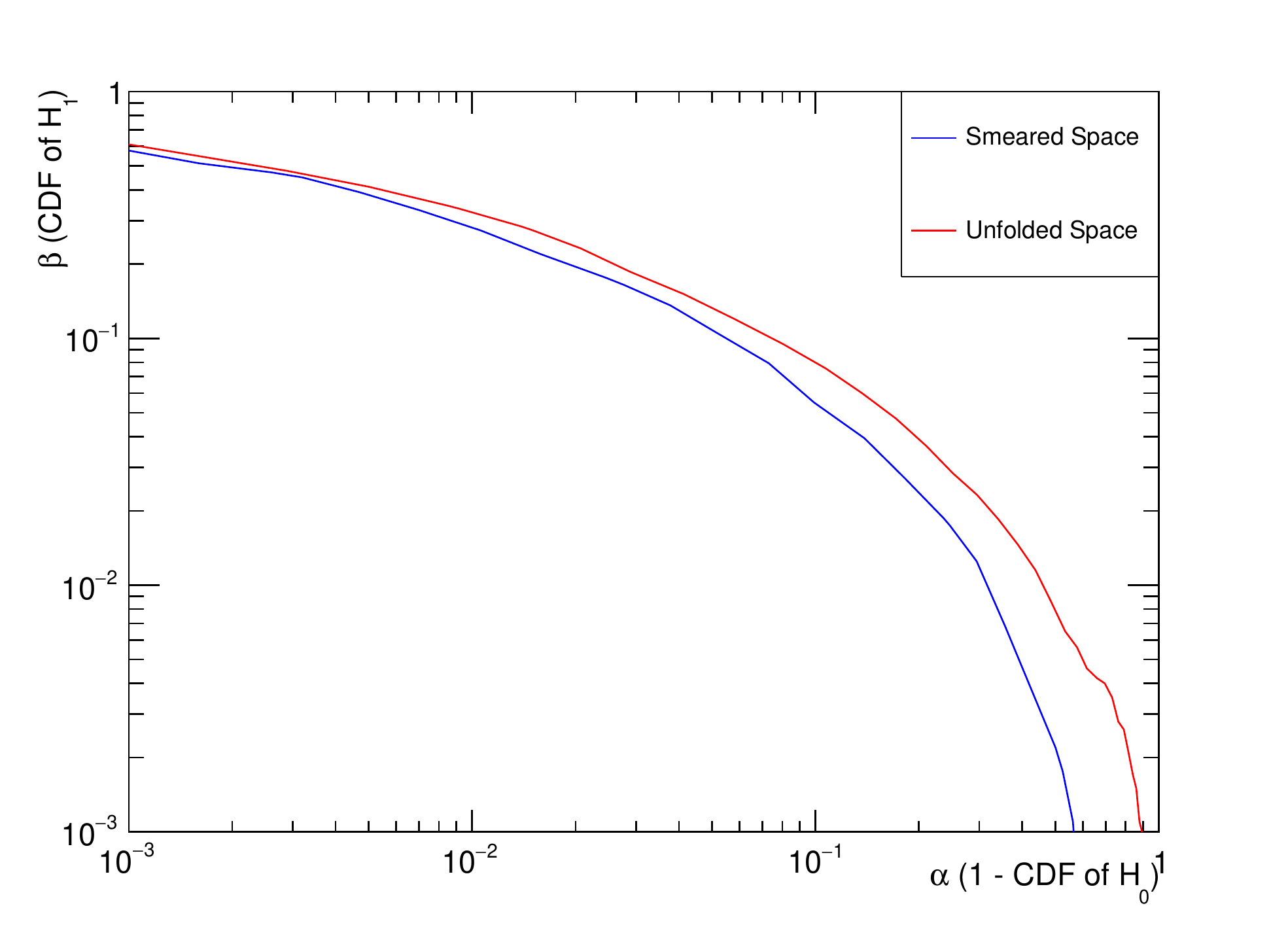}
\caption{The same plot of $\beta$ vs.\ $\alpha$ as in Fig.~\ref{alphabeta}, 
here with logarithmic scale on both axes.
%--------HypothesisComparison\_Invert page 7
%--------HypothesisComparison page 7
}
\label{alphabetaloglog}
\end{center}
\end{figure}

\clearpage
We can investigate the effect of the differences apparent in
Fig.~\ref{deldel} by using the language of Neyman-Pearson hypothesis
testing, in which one rejects $H_0$ if the value of the test statistic
($-2\ln\lambda_{0,1}$ in the smeared space, or $\Delta\chicorr$ in the
unfolded space) is above some critical value~\cite{james}. The Type I
error probability $\alpha$ is the probability of rejecting $H_0$ when
it is true, also known as the ``false positive rate''.  The Type II
error probability $\beta$ is the probability of accepting (not
rejecting) $H_0$ when it is false.  The quantity $1-\beta$ is the {\em
power} of the test, also known as the ``true positive rate''.  The
quantities $\alpha$ and $\beta$ thus follow from the cumulative
distribution functions (CDFs) of histograms of the test statistics.
In classification problems outside HEP is it common to make the ROC
curve of true positive rate vs.\ the false positive rate, as shown in
Fig.~\ref{ROC}.  Figure~\ref{alphabeta} shows the same information in
a plot of $\beta$ vs. $\alpha$, i.e., with the vertical coordinate
inverted compared to the ROC curve.  Figure~\ref{alphabetaloglog} is
the same plot as Fig.~\ref{alphabeta}, with both axes having
logarithmic scale.

The result of this ``bottom line test'' does not appear to be dramatic
in this first example, and appear to be dominated by the difference
between the Poisson-based $-2\ln\lambda_{0,1}$ and $\Delta\chicorr$
already present in the ML unfolding solution, rather than by the
additional differences caused by truncating the EM solution.
Unfortunately no general conclusion can be drawn from this
observation, since as mentioned above the EM unfolding used here
starts from the true distribution as the first estimate.  It is of
course necessary to study other initial estimates.

\clearpage
\subsection{Variation of parameters from the default values}

With the above plots forming a baseline, we can ask how some of the
above plots vary as we change the parameters in Table~\ref{baseline}.

\begin{figure}
\begin{center}
\includegraphics[width=0.49\textwidth]{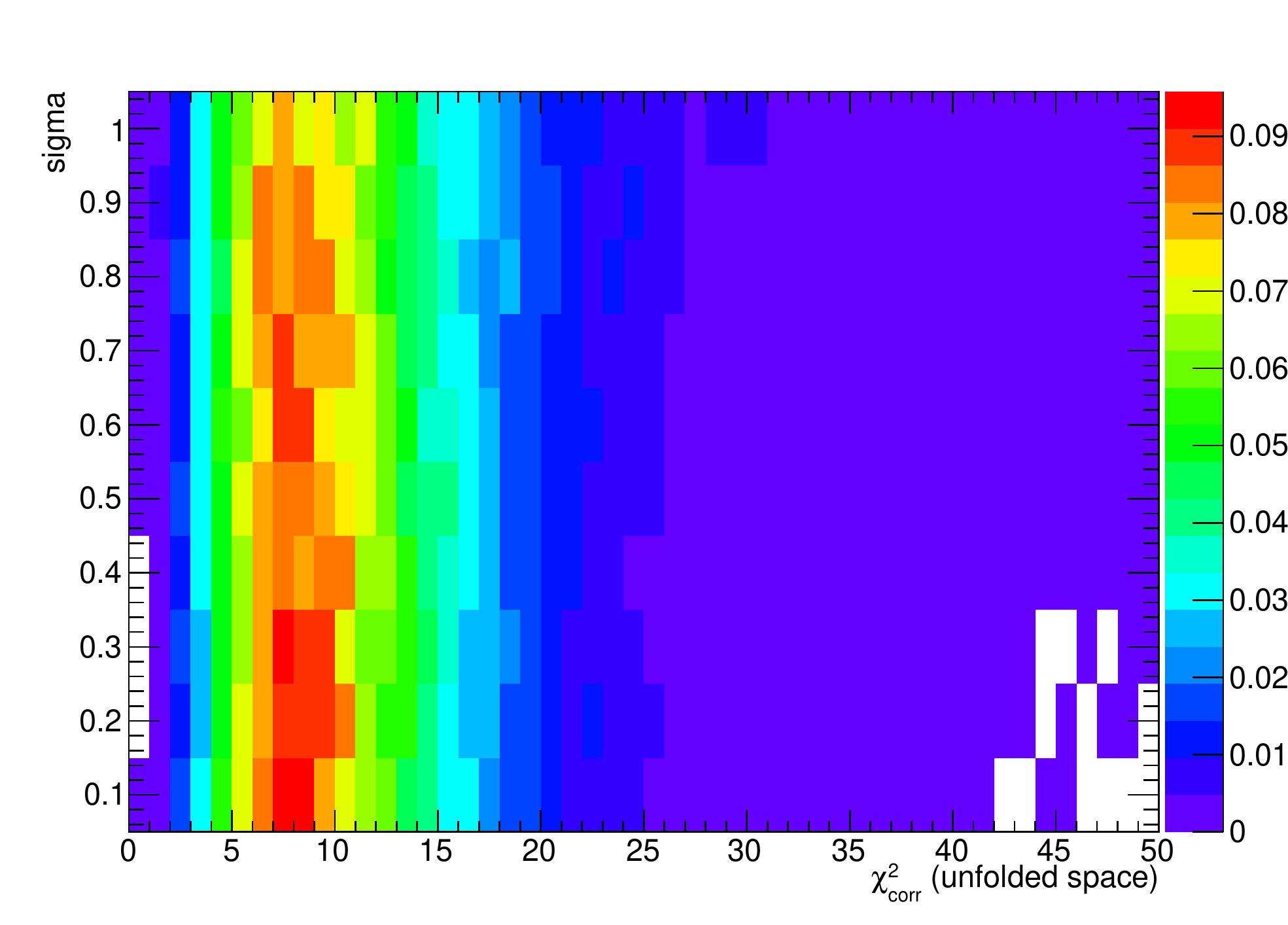}
\includegraphics[width=0.49\textwidth]{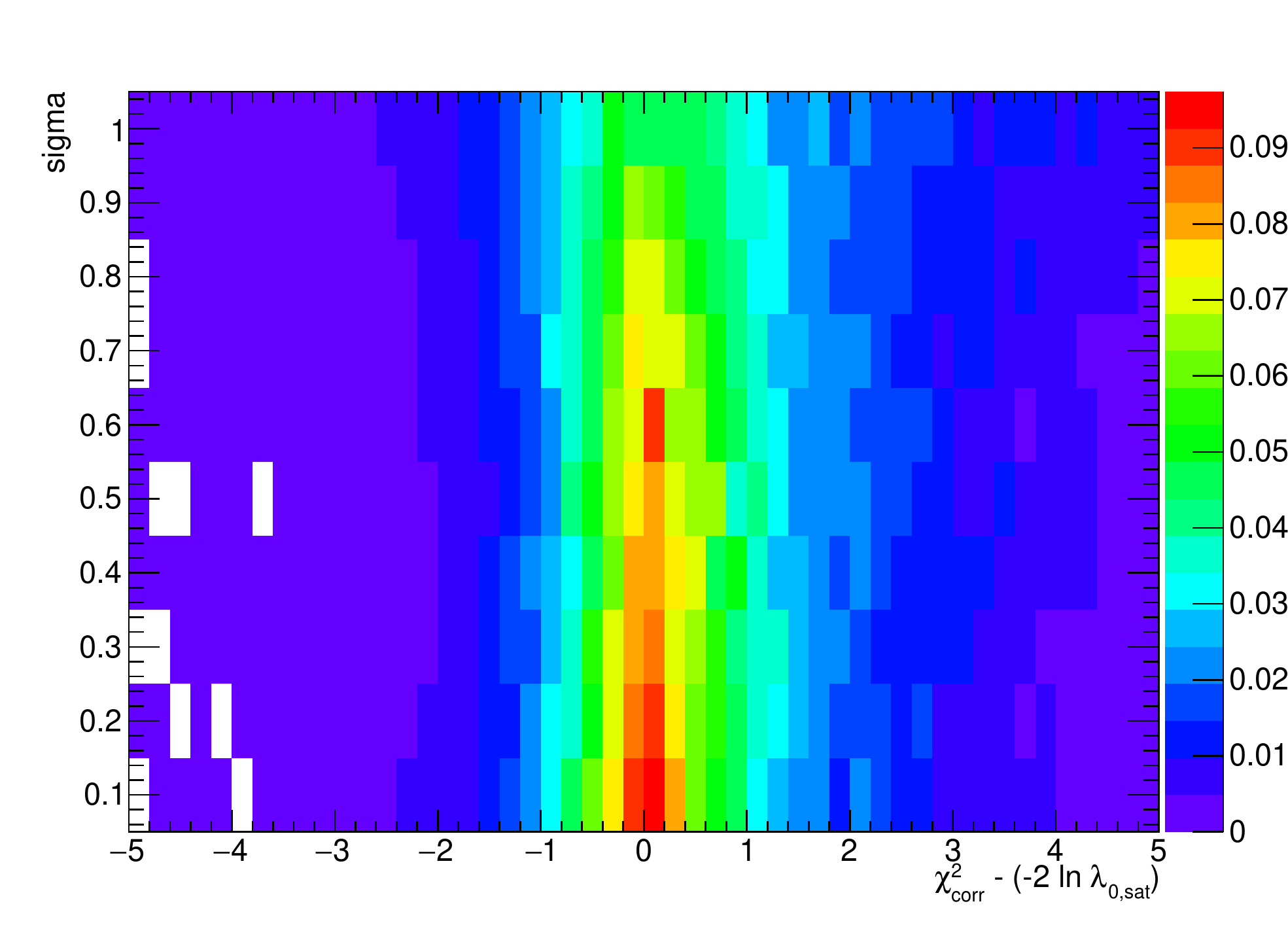}
\caption{For iterative EM unfolding, variation of GOF results with the
  Gaussian $\sigma$ used in smearing (vertical axis).  The horizontal
  axes are the same as those in the 1D histograms in
  Figs.~\ref{nulgofunfoldedinvert} (top left) and
  \ref{nulgofunfoldedinvert} (bottom), namely $\chicorr$ in the
  unfolded space; and the difference with respect to
  $-2\ln\lambda_{0,{\rm sat}}$ in the smeared space; for GOF tests
  with respect to $H_0$ using events generated under $H_0$.
%----------- OneTruth\_2DChi2Histograms.pdf 3 6 
}
\label{sigmaparam}
\end{center}
\end{figure}

Figure~\ref{sigmaparam} shows, as a function of the Gaussian smearing parameter $\sigma$, 
the variation of the GOF results shown for $\sigma=0.5$ in 1D histograms in 
Figs.~\ref{nulgofunfoldedinvert} (top left)
and \ref{nulgofunfoldedinvert} (bottom).  The events are generated under $H_0$.

\begin{figure}
\begin{center}
\includegraphics[width=0.49\textwidth]{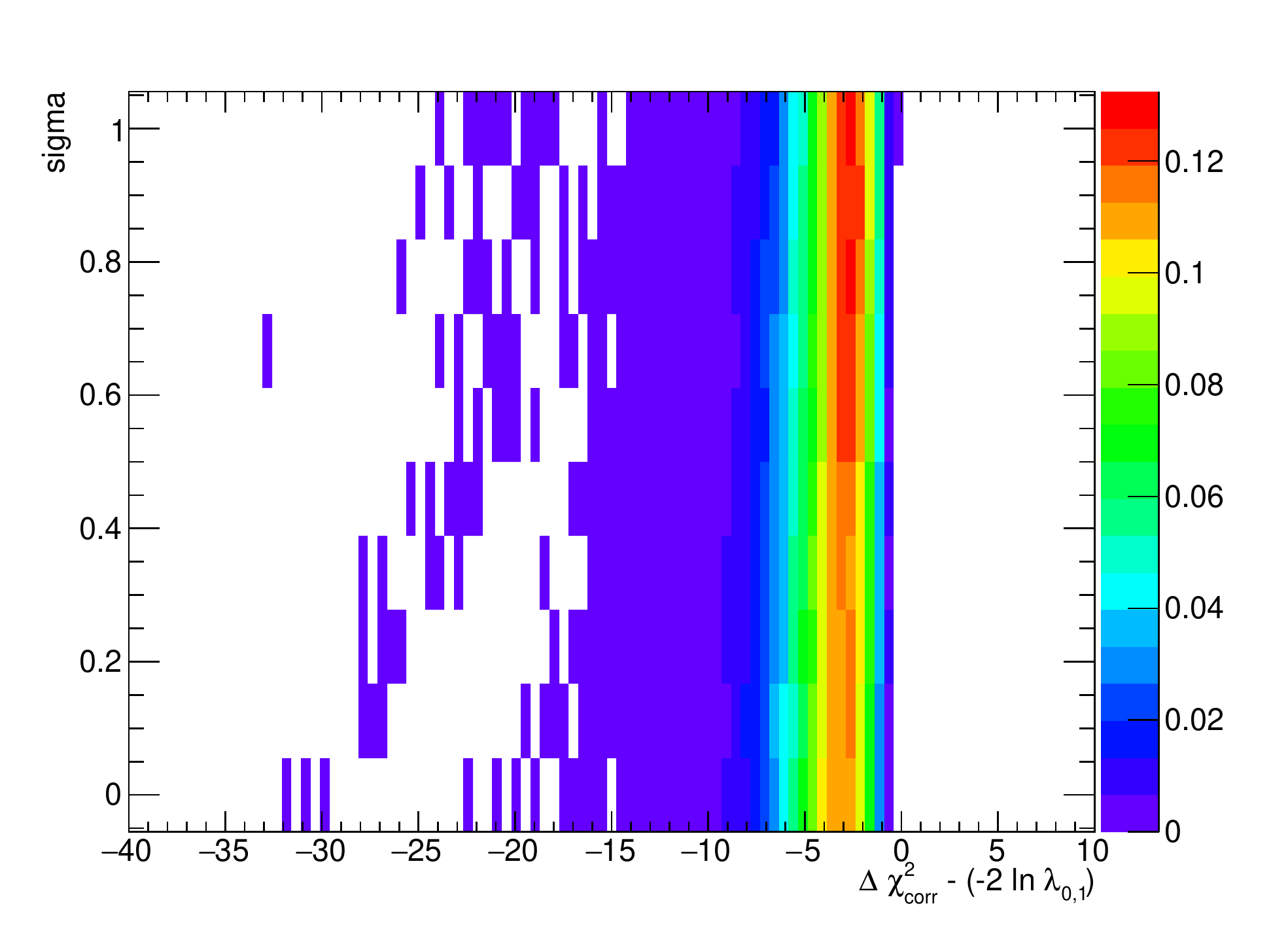}
\includegraphics[width=0.49\textwidth]{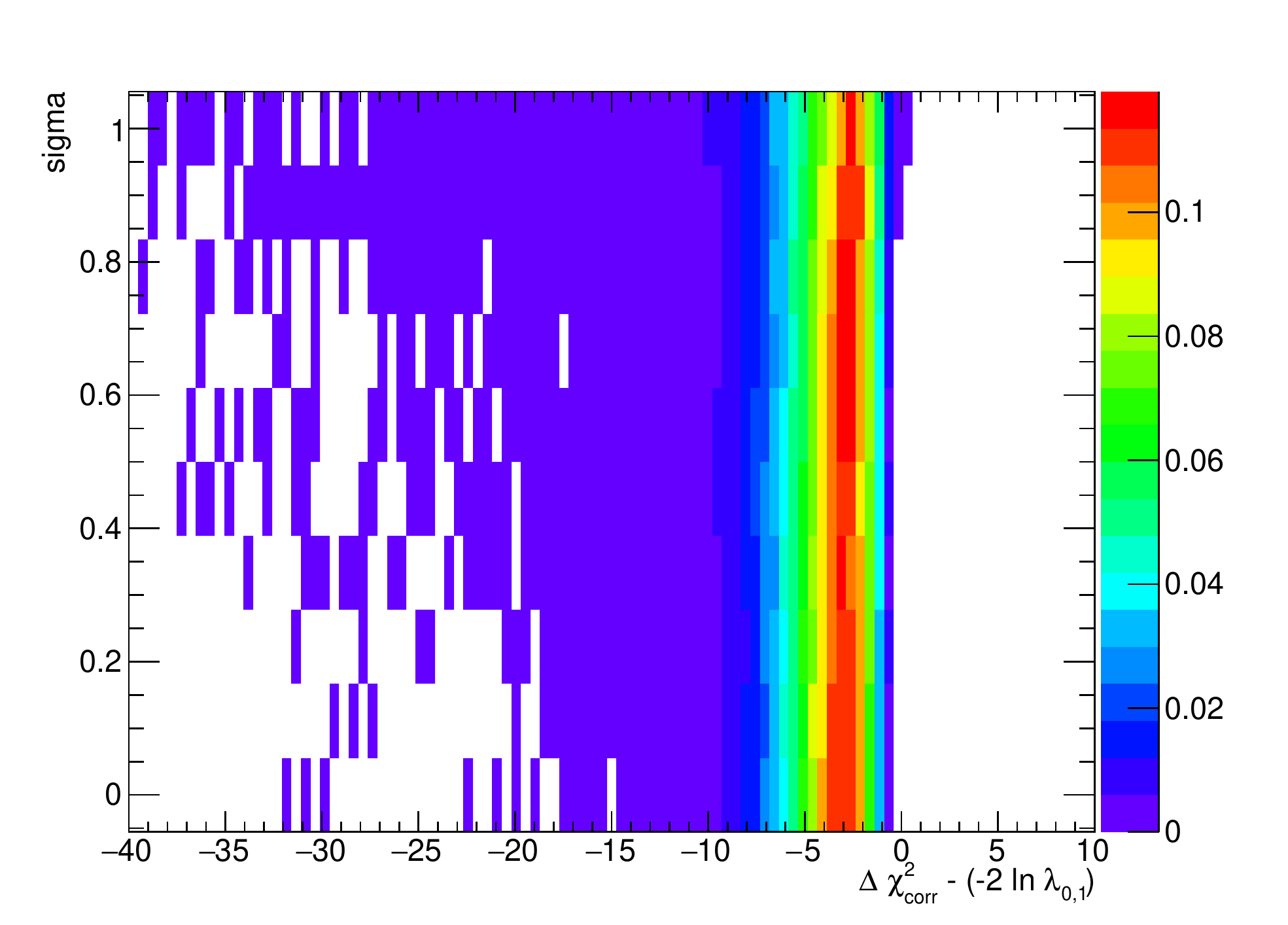}
\caption{(left) For unfolding by ML estimates, variation with the
  Gaussian $\sigma$ used in smearing (vertical axis) of the 1D
  histogram in Fig~\ref{deldel} of the event-by-event difference of
  $-2\ln\lambda_{0,1}$ and $\Delta\chicorr$.  (right) the same
  quantity for iterative EM unfolding.
%---------method\_Invert-param\_sigma-train\_null 17 
%-------method\_Iterative-param\_sigma-train\_null  17
}
\label{deldelsigma}
\end{center}
\end{figure}

%\clearpage
Figure~\ref{deldelsigma} shows the variation of the 1D histogram in
Fig~\ref{deldel} with the Gaussian $\sigma$ used in smearing, for both
ML and EM unfolding.

\begin{figure}
\begin{center}
\includegraphics[width=0.49\textwidth]{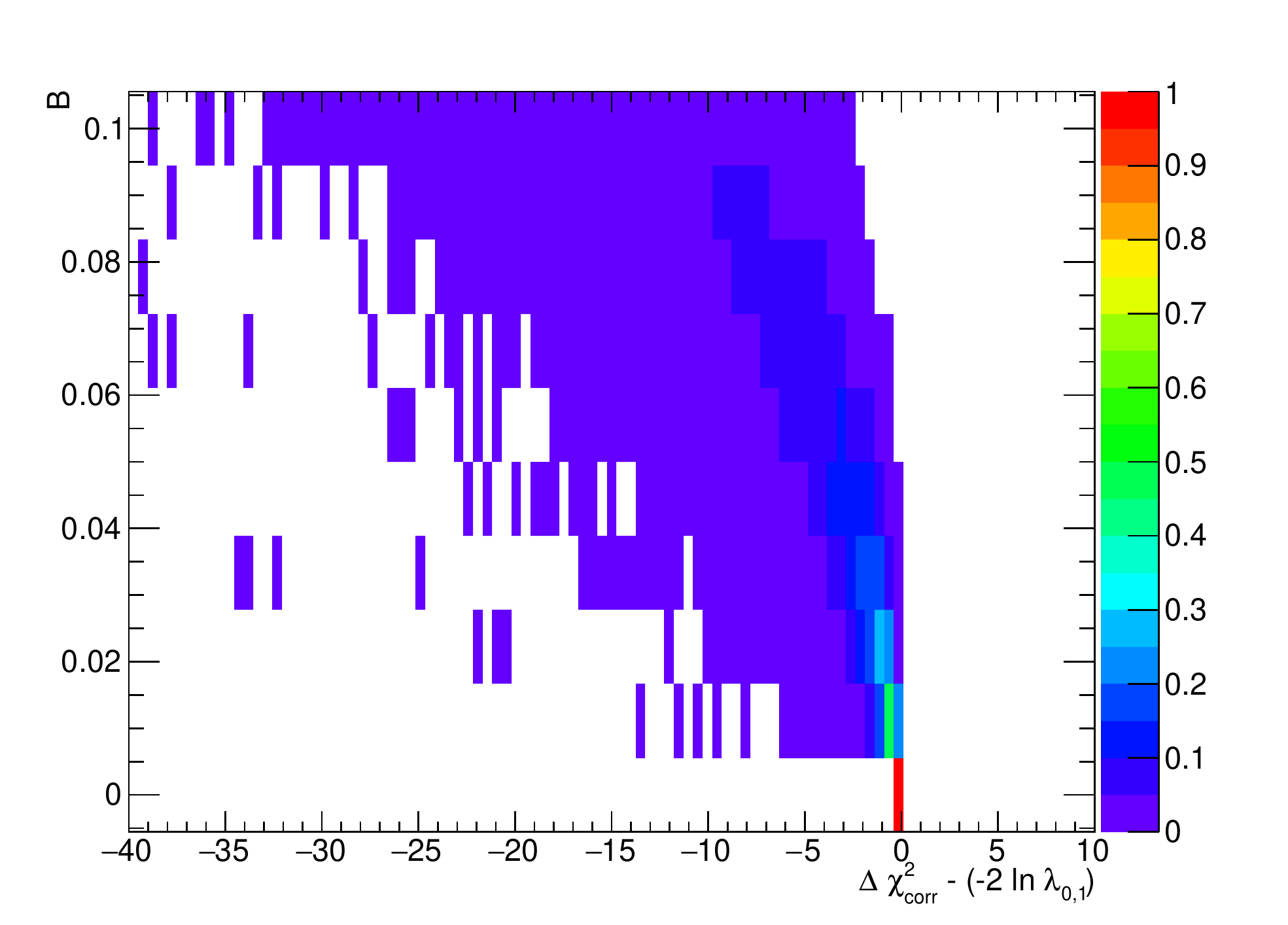}
\includegraphics[width=0.49\textwidth]{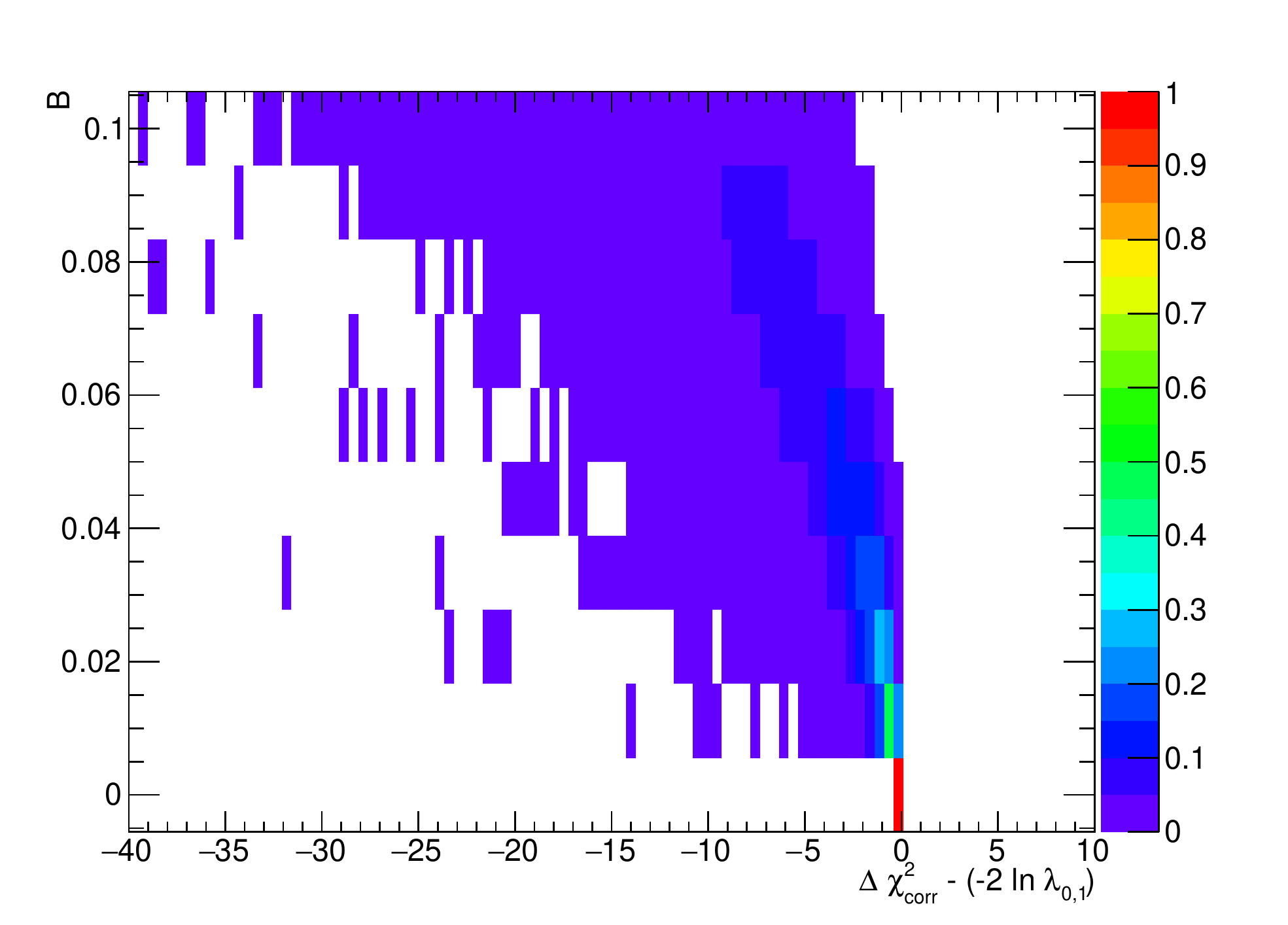}
\caption{The result of the bottom-line test of Fig.~\ref{deldel} as a
  function of the amplitude $B$ of the extra term in $\ptrueone$ in
  Eqn.~\ref{altp}, for (left) $R$ derived from $H_0$ and (right) $R$
  derived from $H_1$; for ML unfolding.
  %-------method\_Invert-param\_f-train\_null 17 
  %-------method\_Invert-param\_f-train\_alt 17 
}
\label{deldelB}
\end{center}
\end{figure}

\begin{figure}
\begin{center}
\includegraphics[width=0.49\textwidth]{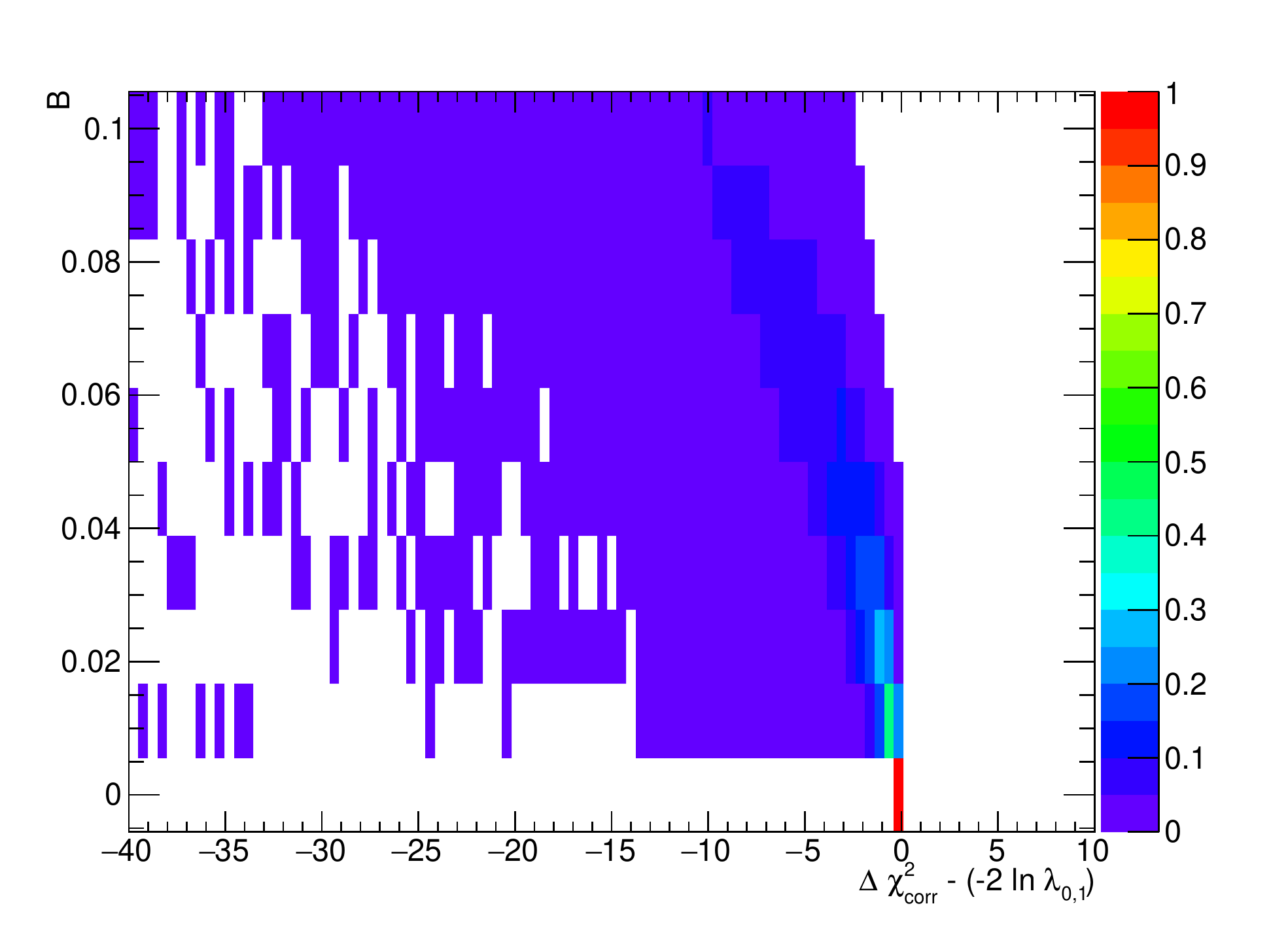}
\includegraphics[width=0.49\textwidth]{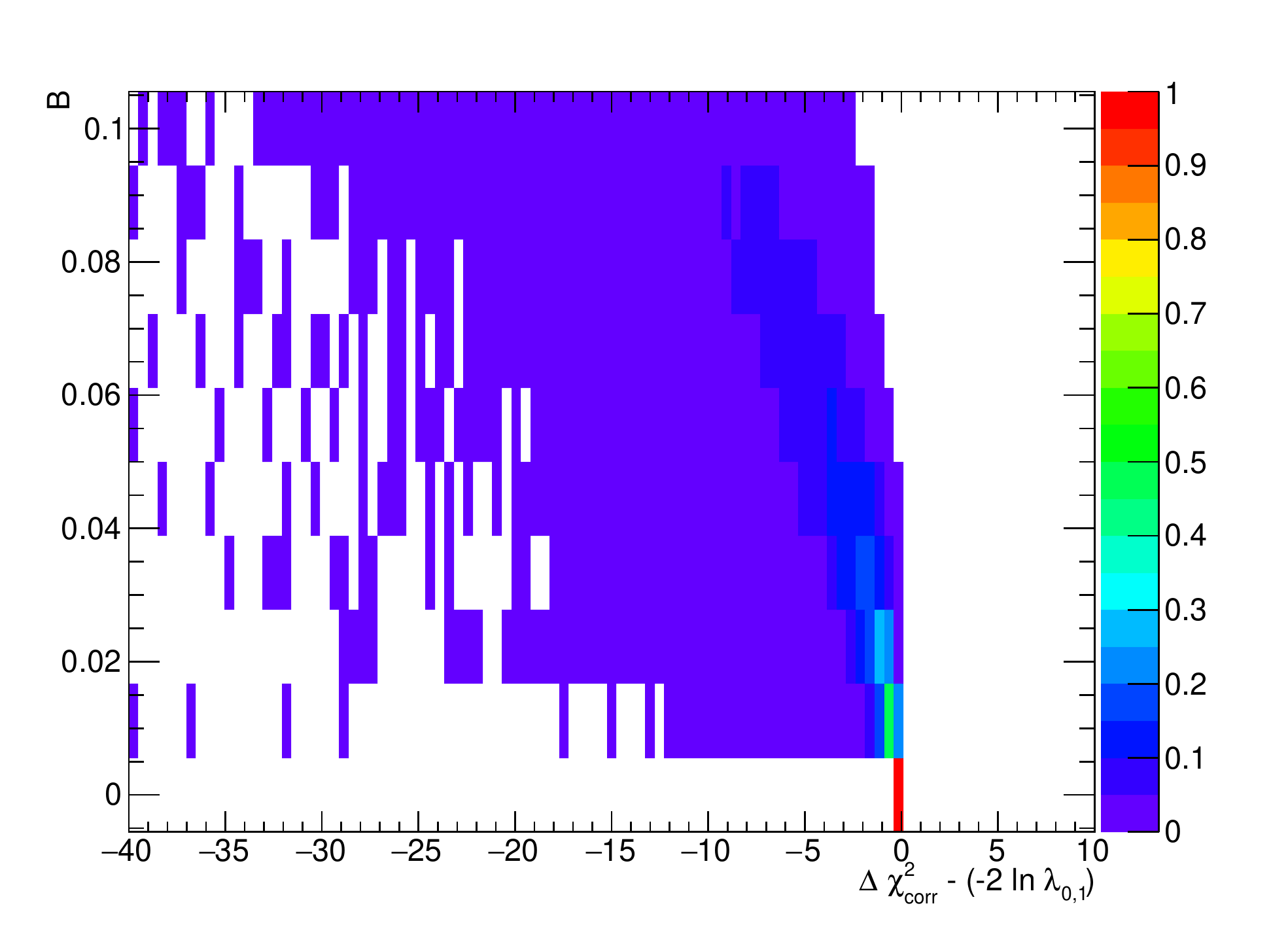}
\caption{The same as Fig.~\ref{deldelB}, for iterative EM unfolding.
%---------method\_Iterative-param\_f-train\_null 17
%--------method\_Iterative-param\_f-train\_alt 17
}
\label{deldelBiter}
\end{center}
\end{figure}

%\clearpage
Figures~\ref{deldelB} and \ref{deldelBiter} show, for ML and EM
unfolding respectively, the result of the bottom-line test of
Fig.~\ref{deldel} as a function of the amplitude $B$ of the extra term
in $\ptrueone$ in Eqn.~\ref{altp}.

\begin{figure}
\begin{center}
\includegraphics[width=0.49\textwidth]{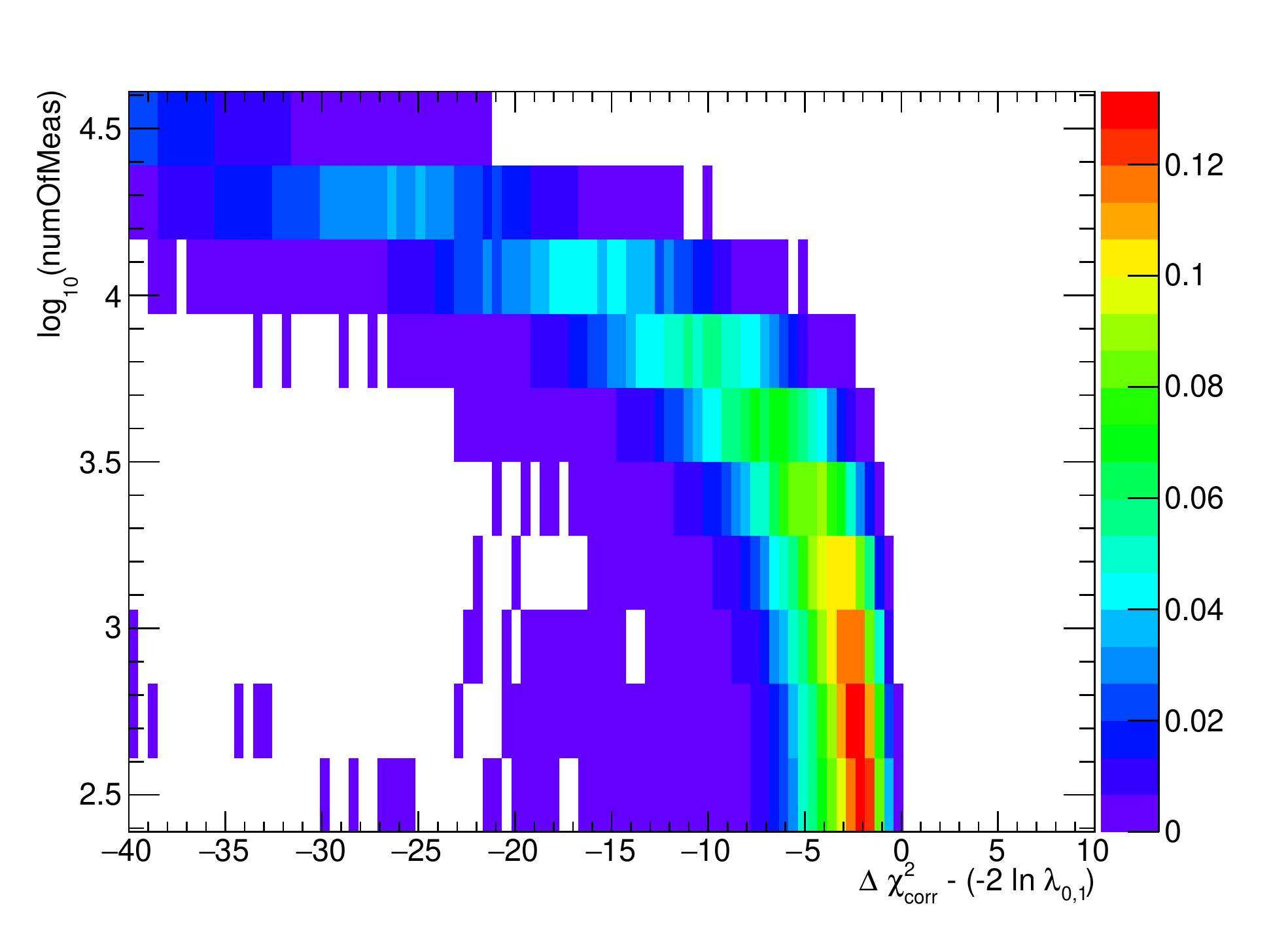}
\includegraphics[width=0.49\textwidth]{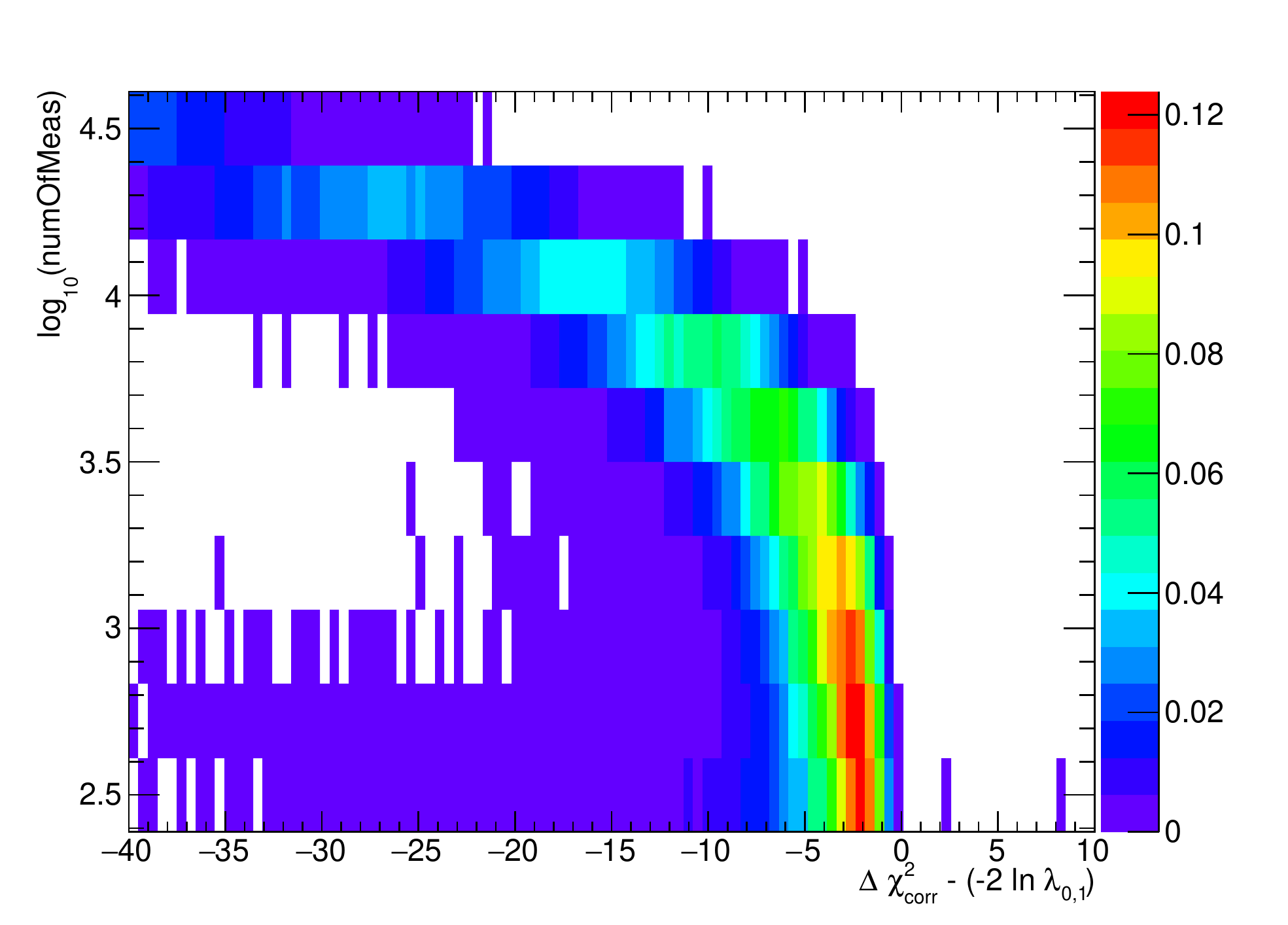}
\caption{The result of the bottom-line test of Fig.~\ref{deldel} as a
  function of the number of events on the histogram of $\vec n$, for
  (left) ML unfolding and (right) iterative EM unfolding.
%----------method\_Invert-param\_numOfMeas\_log-train\_null 17
%---------method\_Iterative-param\_numOfMeas\_log-train\_null 17
}
\label{deldelnummeas}
\end{center}
\end{figure}

%\clearpage
Figure~\ref{deldelnummeas} shows, for ML and EM unfolding, the result
of the bottom-line test of Fig.~\ref{deldel} as a function of the
mean number of events in the histogram of $\vec n$.

\begin{figure}
\begin{center}
\includegraphics[width=0.49\textwidth]{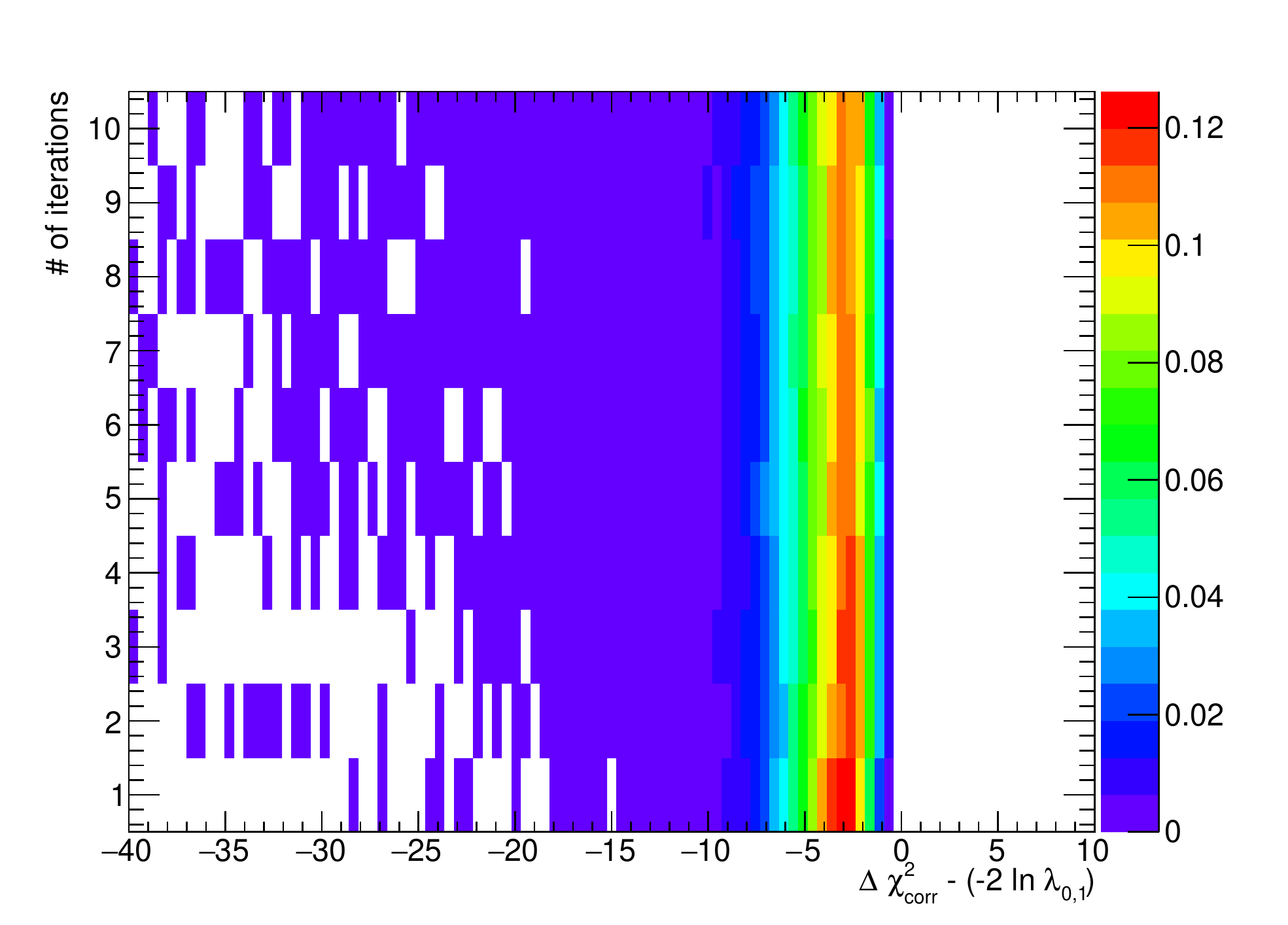}
\includegraphics[width=0.49\textwidth]{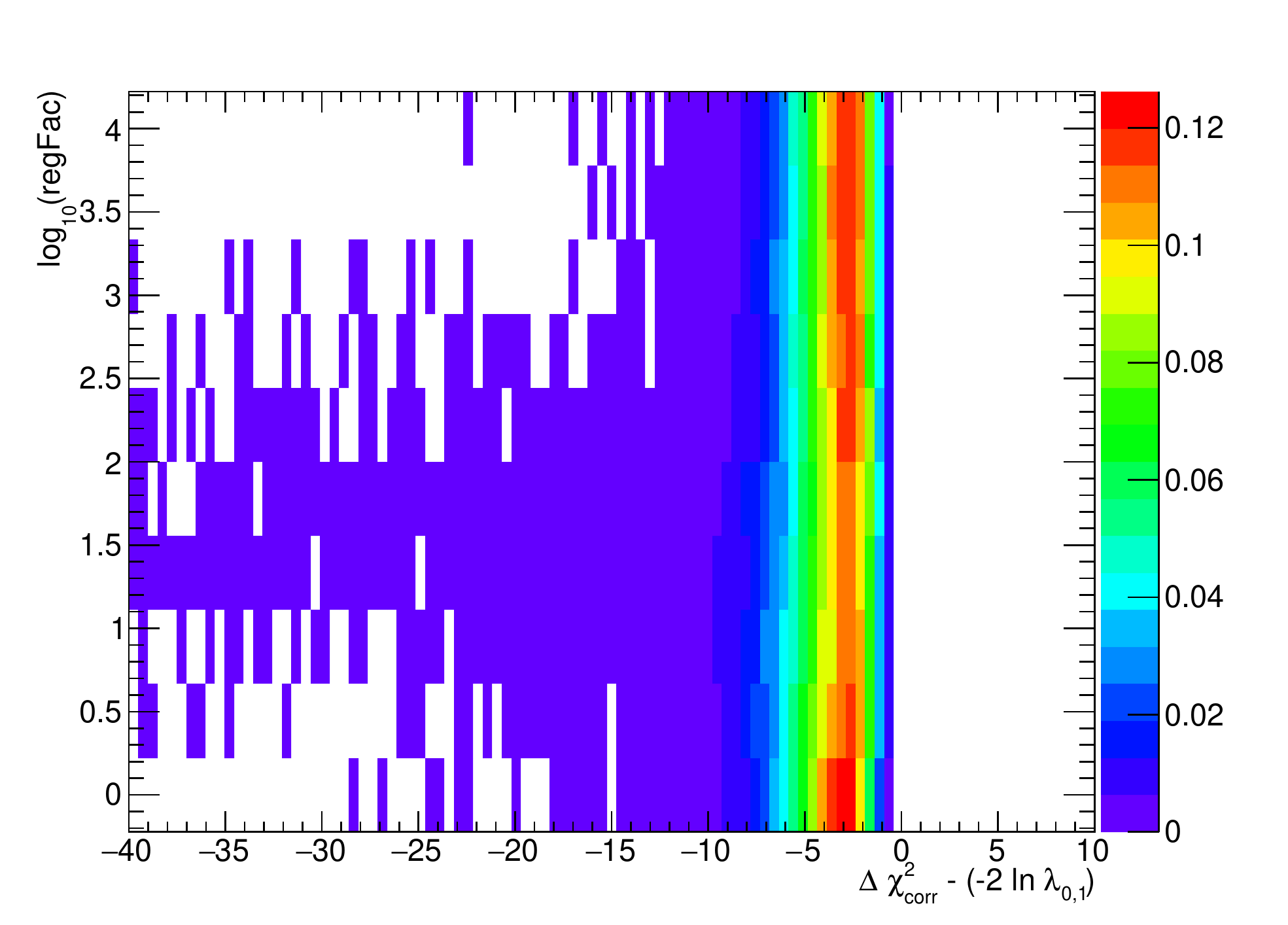}
\caption{For EM iterative unfolding, the result of the bottom-line
  test of Fig.~\ref{deldel} as a function of number of iterations in
  (left) linear vertical scale and (right) logarithmic vertical scale.
%---------method\_Iterative-param\_regFac-train\_null 11
%-------method\_Iterative-param\_regFac\_log-train\_null 11
}
\label{deldelreg}
\end{center}
\end{figure}

Figure~\ref{deldelreg} shows, for iterative EM unfolding, the result
of the bottom-line test of Fig.~\ref{deldel} as a function of the
number of iterations.

\clearpage
\section{Discussion}
This note illustrates in detail some of the differences that can arise
with respect to the smeared space when testing hypotheses in the
unfolded space.  As the note focuses on a particularly simple
hypotheses test, and looks only at the ML and EM solutions, no general
conclusions can be drawn, apart from claiming the potential usefulness
of the ``bottom line tests''. Even within the limitations of the
RooUnfold software used here (in particular that the initial estimate
for iterating is the presumed truth), we see indications of dangers of
testing hypotheses after unfolding.  Perhaps the most interesting
thing to note thus far is that unfolding by matrix inversion (and
hence no regularization) yields, in the implementation studied here, a
generalized $\Delta\chicorr$ test statistic that is identical to
$\chin$ in the smeared space, which is intrinsically inferior to
$-2\ln\lambda_{0,1}$.  The potentially more important issue of bias
due to regularization affecting the bottom line test remains to be
explored.

Such issues should be kept in mind, even in informal comparisons of
unfolded data to predictions from theory.  For quantitative comparison
(including the presumed use of unfolded results to evaluate
predictions in the future from theory), we believe that extreme
caution should be exercised, including performing the
bottom-line-tests with various departures from expectations.  This
applies to both GOF tests of a single hypothesis, and comparisons of
multiple hypotheses.

More work is needed in order to gain experience regarding what sort of
unfolding problems and unfolding methods yield results that give
reasonable performance under the bottom-line-test, and which cases
lead to bad failures.  As often suggested, reporting the response
matrix $R$ along with the smeared data can facilitate comparisons with
future theories in the folded space, in spite of the dependence of $R$
on the true pdfs.

\section*{Acknowledgments}
We are grateful to Pengcheng Pan, Yan Ru Pei, Ni Zhang, and Renyuan
Zhang for assistance in the early stages of this study.  RC thanks the
CMS Statistics Committee and G\"unter Zech for helpful discussions
regarding the bottom-line test.  This work was partially supported by
the U.S.\ Department of Energy under Award Number {DE}--{SC}0009937.

\end{document}